\documentclass{emulateapj}
\usepackage{amsmath}
\usepackage{amssymb}
\usepackage{natbib}
\usepackage{graphicx}
\usepackage{color}

\shorttitle{The Dark Halo -- Spheroid Conspiracy in Elliptical Galaxies}
\shortauthors{Remus et al.}

\begin{document}
\submitted{Published by the Astrophysical Journal (ApJ, 766, 71)}

\title{The Dark Halo -- Spheroid Conspiracy and the Origin of Elliptical Galaxies}

\author{Rhea-Silvia Remus$^1$$^2$, Andreas Burkert$^1$$^2$, Klaus Dolag$^1$$^4$, Peter H. Johansson$^3$,\\ Thorsten Naab$^4$, Ludwig Oser$^4$, and Jens Thomas$^2$}
\affil{$^1$ Universit\"ats-Sternwarte M\"unchen, Scheinerstr.\ 1, D-81679 M\"unchen, Germany\\
$^2$ Max-Planck-Institute for Extraterrestrial Physics, P.O. Box 1312, D-85748 Garching, Germany\\
$^3$ Department of Physics, University of Helsinki, Gustaf H\"allstr\"omin katu 2a, FI-00014 Helsinki, Finland\\
$^4$ Max-Planck-Institute for Astrophysics, Karl-Schwarzschild-Str.\ 1, D-85748 Garching, Germany \\ \texttt{rhea@usm.lmu.de} \\}

\begin{abstract}
Dynamical modeling and strong lensing data indicate that the total density profiles of early-type galaxies are close to isothermal, i.e., $\rho_{\rm tot} \propto r^{\gamma}$ with $\gamma \approx -2$.  
To understand the origin of this universal slope we study a set of simulated spheroids formed in isolated binary mergers as well as the formation within the cosmological framework.
The total stellar plus dark matter density profiles can always be described by a power law with an index of $\gamma \approx -2.1$ with a tendency toward steeper slopes for more compact, lower-mass ellipticals.
In the binary mergers the amount of gas involved in the merger determines the precise steepness of the slope.
This agrees with results from the cosmological simulations where ellipticals with steeper slopes have a higher fraction of stars formed in situ.
Each gas-poor merger event evolves the slope toward $\gamma \sim -2$, once this slope is reached further merger events do not change it anymore.
All our ellipticals have flat intrinsic combined stellar and dark matter velocity dispersion profiles.
We conclude that flat velocity dispersion profiles and total density distributions with a slope of $\gamma \sim -2$ for the combined system of stars and dark matter act as a natural attractor.
The variety of complex formation histories as present in cosmological simulations, including major as well as minor merger events, is essential to generate the full range of observed density slopes seen for present-day elliptical galaxies.
\end{abstract}


\section{Introduction}
Early-type galaxies are among the most massive and prominent galaxies of the universe, and thus their origin and formation history has been the focus of many observational studies as well as simulations.
Since \citet{toomre:1977egsp.conf..401T} proposed that elliptical galaxies can be produced by a merger between two spiral galaxies, several simulations have studied this mechanism in detail, for example, \citet{white:1978MNRAS.184..185W,white:1979ApJ...229L...9W,white:1979MNRAS.189..831W,gerhard:1981MNRAS.197..179G,negroponte:1983MNRAS.205.1009N,barnes:1988ApJ...331..699B}, including various gas fractions \citep{hernquist:1989Natur.340..687H,barnes:1996ApJ...471..115B,naab:2006MNRAS.372..839N,novak:2012MNRAS.424..635N} and black hole (BH) physics \citep{springel:2005ApJ...620L..79S,johansson:2009ApJ...707L.184J,johansson:2009ApJ...690..802J}.

Although major merger simulations of spiral galaxies could explain many observed properties of ellipticals, several problems were also identified.
A number of studies have indicated that observations of massive, slowly rotating and spheroidal ellipticals cannot be explained by wet or dry major mergers of disk galaxies \citep{naab:2003ApJ...597..893N,cox:2006ApJ...650..791C,burkert:2008ApJ...685..897B,bois:2010MNRAS.406.2405B,bois:2011MNRAS.416.1654B}.
Multiple early-type mergers would be needed in order to form the most massive ellipticals with masses that exceed the mass of spiral galaxies \citep{naab:2006ApJ...636L..81N,gonzalez:2006MNRAS.372L..78G,khochfar:2006ApJ...648L..21K}.
\citet{genel:2010ApJ...719..229G} however demonstrated that the fraction of observed major early-type mergers is not high enough to explain the number of observed giant elliptical galaxies.
An additional complication is that the high metallicity and old ages observed in present-day massive early-type galaxies cannot be explained by binary mergers between two typical present-day spiral galaxies or their progenitors (\citealp{naab:2009ApJ...690.1452N}, see however \citealp{hopkins:2008ApJ...679..156H}).

Simulations of the formation of elliptical galaxies in a full cosmological context result in a different overall picture (e.g., \citealp{meza:2003ApJ...590..619M,naab:2007ApJ...658..710N,naab:2009ApJ...699L.178N,gonzales:2009A&A...497...35G,feldmann:2010ApJ...709..218F,feldmann:2011ApJ...736...88F,oser:2010ApJ...725.2312O,oser:2012ApJ...744...63O,johansson:2012ApJ...754..115J,lackner:2012MNRAS.425..641L}), verifying the idea that elliptical galaxies can be produced by multiple minor mergers, a much more likely evolution of events in the lifetime of a galaxy (see also \citealp{bournaud:2007A&A...476.1179B}). 
Mass accretion histories from cosmological simulations (e.g., \citealp{gao:2004ApJ...614...17G,fakhouri:2008MNRAS.386..577F,fakhouri:2009MNRAS.394.1825F}) have shown that the most likely process for forming elliptical galaxies is a mixture of both scenarios: most massive halos go through an early dissipative phase of fast accretion, including major merger events, followed by a phase of stellar accretion, during which all kind of mergers and even smooth accretion can be the dominant growth mechanism (e.g., \citealp{oser:2010ApJ...725.2312O}).

There have been several approaches to connect the observed elliptical properties to the different formation scenarios to investigate if we can explain observed features by the evolution history of individual ellipticals.
For example, \citet{naab:1999ApJ...523L.133N,bendo:2000MNRAS.316..315B,naab:2003ApJ...597..893N,jesseit:2005MNRAS.360.1185J,cox:2006ApJ...650..791C} and \citet{gonzales:2009A&A...497...35G} studied the kinematics and photometric shapes of simulated spheroidals in order to understand the origin of the shape of the galaxy, the origin of boxy and disky isophotes, and the connections between the kinematic properties and the gas fraction of the merger event, \citet{naab:2006MNRAS.372..839N} showed that the line-of-sight velocity dispersions are strongly influenced by gas (see also \citealp{hoffman:2009ApJ...705..920H,hoffman:2010ApJ...723..818H}), while \citet{qu:2010A&A...515A..11Q} and \citet{diMatteo:2009A&A...501L...9D} showed that multiple minor mergers slow down the rotation of an elliptical, while major mergers can speed them up.

Recent results from the Atlas3D survey \citep{cappellari:2011MNRAS.413..813C}, which uses integral field spectroscopy to study the properties of early-type galaxies, have revealed that the vast majority of all early-type galaxies (82\%--86\%) in their volume-limited local galaxy sample is fast rotating, while only very few early-type galaxies are classified as slow or even non-rotating \citep{emsellem:2011MNRAS.414..888E}.
A significant fraction of the slow rotating ellipticals show special features such as a counter-rotating core.
The non-rotating early types are usually found in highly overdense environments \citep{krajnovic:2011MNRAS.414.2923K}. 
\citet{bois:2011MNRAS.416.1654B} showed, comparing simulations of isolated merger events to the Atlas3D results, that major mergers can reproduce all kinds of elliptical galaxies that are fast rotating as well as population of flat, slow rotating elliptical galaxies with kinematically distinct cores.
They fail, however, to reproduce very round, massive, slow, or even non-rotating ellipticals that typically live in very dense environments (see also \citealp{burkert:2008ApJ...685..897B}), supporting the idea that the round slow rotators and the fast rotating systems actually are two distinct families of early-type galaxies with different formation histories.

Elliptical galaxies consist mainly of a stellar and a dark matter component that might provide interesting information about their formation history.
For example, it is long known from observations that there exists a dark halo--disk conspiracy for spiral galaxies, i.e., the rotation curves of spiral galaxies are flat \citep{einasto:1974Natur.250..309E,faber:1979ARA&A..17..135F}.
Detailed dynamical modeling, for example by \citet{kronawitter:2000A&AS..144...53K} and \citet{gerhard:2001AJ....121.1936G}, has demonstrated that a similar conspiracy might also exist between the dark halo and the spheroid of massive, slow rotating elliptical galaxies.

Further studies, using planetary nebulae as tracers for the outer dark halo, however have revealed a bimodality in the kinematic structures of the outer regions of elliptical galaxies: some lower-mass ellipticals show declining velocity dispersion profiles, similar to a Keplerian mass distribution, indicating a constant mass-to-light ratio and thus a shallow dark matter halo \citep{mendez:2001ApJ...563..135M,romanowsky:2003Sci...301.1696R}, while especially more massive, slow rotating galaxies show flat dispersion profiles as expected for a classical, extended dark matter halo \citep{napolitano:2001A&A...377..784N,peng:2004ApJ...602..685P}.
The origin of this bimodality has been discussed, suggesting for example that extremely elongated orbits of stars and planetary nebulae in the outskirts of low-mass ellipticals can mask even a massive dark matter halo \citep{dekel:2005Natur.437..707D}. 
A recent study by \citet{deason:2012ApJ...748....2D} investigated dark matter fractions and density and velocity dispersion profiles of 15 elliptical galaxies out to 5 effective radii, using planetary nebulae and globular clusters as tracers.

A more detailed look at the de-projected dark matter component of elliptical galaxies inside the half-light radius is provided by \citet{thomas:2007MNRAS.382..657T,thomas:2009ApJ...691..770T}, using Schwarzschild modeling of the observational kinematic and photometric data of early-type galaxies in the Coma cluster.
They show that the best-fitting models contain 10 to 50 percent dark matter within the half-light radius. Furthermore, their models without dark matter halos are not able to fit the observations.

Another approach to study the dark matter content of elliptical galaxies comes from strong lensing.
\citet{auger:2010ApJ...724..511A}, studying 73 early-type galaxies from the SLACS survey, found that the total dark matter plus stellar density profiles of these ellipticals can be described on average by a power law with a slope of $-2.078$, with the steepness of the individual slopes correlating with the effective radius and the central density of the stellar component. 
A similar result was found by \citet{barnabe:2011MNRAS.415.2215B} for 16 early-type galaxies from SLACS, where they combined the constraints from gravitational lensing with stellar kinematics.
Further studies by \citet{ruff:2011ApJ...727...96R} and \citet{bolton:2012ApJ...757...82B} revealed that this is also the case for strong-lensing early-type galaxies at higher redshifts: \citet{ruff:2011ApJ...727...96R} found that their 11 early-type galaxies in a redshift range of $0.2 \lesssim z \lesssim 0.65$ can be fit on average by a slope of $-2.16$, while \citet{bolton:2012ApJ...757...82B} found for their sample of 79 early-type galaxies in the redshift range of $0.1 \lesssim z \lesssim 0.6$ a density slope of on average $-2.11$.
Both report a slight trend to flatter slopes at higher redshifts.

An additional method to measure the dark matter content of early-type galaxies is weak lensing, which enables measurements to much larger radii than strong lensing.
\citet{gavazzi:2007ApJ...667..176G} showed for 22 early-type galaxies from weak lensing that their total density profiles can be described by a power law with a slope of approximately $-2$ for radii as large as 300~kpc, which is up to 100 effective radii.

In a recent study, \citet{lyskova:2012MNRAS.423.1813L} investigated a sample of cosmological simulations of massive galaxies presented by \citet{oser:2010ApJ...725.2312O} and found a remarkable uniformity of the present-day isothermal total stellar mass profiles in good agreement with lensing results. In this work we investigate the structure of dark matter halos (density distribution and kinematics) around simulated ellipticals in more detail and relate it to the formation mechanisms and histories of elliptical galaxies.
Using a variety of different simulations allows us to study the effects of the multiple merger evolution in comparison to the major merger scenario as well as the influence of the environment on the resulting dynamical profiles.

The paper is structured as follows: In Section~\ref{sec:simulations} we describe the simulations used for this work. Section~\ref{sec:analysis} explains how we selected and analyzed the elliptical galaxies from the simulations. In Section~\ref{sec:conclusions} we discuss our results.

\section{Numerical Simulations}\label{sec:simulations}
We use ellipticals formed in different scenarios: Isolated binary merger simulations with controlled initial conditions (hereafter Binary Merger), hydrodynamical cosmological zoom-in simulations (hereafter CosmoZoom) and large-scale hydrodynamical cosmological simulations (hereafter Magneticum Pathfinder).
All simulations were performed using extended versions of the parallel TreePM-SPH-code GADGET-2 \citep{springel:2005MNRAS.364.1105S} and include simplified merger scenarios as well as simulations within the cosmological framework, where halos merge in a hierarchical fashion.

GADGET-2 is based on an entropy-conserving formulation of SPH \citep{springel:2002MNRAS.333..649S} and in its standard version includes radiative cooling for a primordial mixture of hydrogen and helium \citep{katz:1996ApJS..105...19K}.
Star formation and the associated supernova feedback is included using a sub-resolution model \citep{springel:2003MNRAS.339..289S} and assumes a Salpeter initial mass function (IMF) \citep{salpeter:1955ApJ...121..161S}.
In this model the ISM is treated as a two-phase medium \citep{mcKee:1977ApJ...218..148M,efstathiou:2000MNRAS.317..697E,johansson:2006MNRAS.371.1519J} in which cold clouds are embedded in a tenuous hot gas at pressure equilibrium.
In the cosmological simulations this also includes heating by the time dependent but spatially uniform UV background \citep{haardt:1996ApJ...461...20H}.

The Magneticum Pathfinder simulations also follow the pattern of metal production from the past history of cosmic star formation \citep{tornatore:2004MNRAS.349L..19T,tornatore:2007MNRAS.382.1050T}.
This is done by computing the contributions from both Type~II and Type~Ia supernovae, and energy feedback and metals are released gradually in time, with the appropriate lifetimes of the different stellar populations.
This treatment also includes, in a self-consistent way, the dependence of the gas cooling on the local metallicity.
The feedback scheme in this case assumes a Chabrier IMF \citep{chabrier:2003PASP..115..763C}.
It additionally includes kinetic feedback mimicking the effect of supernova winds \citep{springel:2003MNRAS.339..289S} with its parameters fixed to a wind velocity of $\approx 250 \rm ~km\,s^{-1}$. 

Black Holes are included in the binary merger simulations and the Magneticum Pathfinder simulation as sink particles.
BH feedback is taken into account according to the model from \citet{springel:2005MNRAS.361..776S}, where the BH sink particle accretes gas from the surrounding medium according to a Bondi-Hoyle accretion model, limited to the Eddington limit, and gives back thermal feedback to the surrounding medium. 
Two BHs are assumed to merge instantly as soon as they enter each other's smoothing length and their relative velocity is below the local sound speed.

\subsection{Binary Merger Simulations}
The classical formation scenario for an elliptical galaxy is the major merger scenario, where we artificially set up two spiral galaxies and collide them on specified orbits.
We analyze a set of 10 high resolution ellipticals formed in such a major merger scenario, which we will refer to as Binary Ellipticals hereafter.
For a detailed description of the simulations used in this work, especially the details of the setup of the progenitor spirals, see \citet{johansson:2009ApJ...707L.184J} and \citet{johansson:2009ApJ...690..802J}.
Our sample of simulations consists of 4 spiral-spiral mergers with a mass ratio of 1:1 for the progenitor galaxies, 5 spiral-spiral mergers with a mass ratio of 3:1 and one mixed merger.
For the mixed merger we collide a spiral galaxy with an elliptical which is a remnant of a 3:1 spiral-spiral merger itself.

The galaxies were set up following the method presented in \citet{springel:2005MNRAS.361..776S}.
The dark halo virial velocity for all primary galaxies except $31 \rm OBH2\_09\_320$ is $v_{\rm vir} = 160 \rm ~km\,s^{-1}$, halo $31 \rm OBH2\_09\_320$ is set up with $v_{\rm vir} = 320 \rm ~km\,s^{-1}$.
Here $v_{\rm vir}$ defines the dark matter virial mass and virial radius of the spiral galaxy:
\begin{flalign}
M_{\rm vir} = \frac{v_{\rm vir}^3}{10 G H_0} \label{eq:virial1} \\
r_{\rm vir} = \frac{v_{\rm vir}}{10 H_0} \label{eq:virial}
\end{flalign}
The disks are set up with different initial gas fractions, $f_{\rm gas} = 0.0$, $f_{\rm gas} = 0.2$ and $f_{\rm gas} = 0.8$, with the rest being disk stars.
Simulation $11 \rm OBHNB0\_13$ is a simulation without a bulge and a gas fraction of $f_{\rm gas} = 0.0$.
All of the simulations, except $31 \rm ASF2\_13$ and $31 \rm ASF2\_13$, also include Black Holes (BH) as sink particles and BH feedback.

We adopt three orbital geometries, G13, G09 and G01, according to \citet{naab:2003ApJ...597..893N}.
Since we were interested in studying the effects of the initial gas fraction and the merger ratios on the merger remnant, we choose for the majority of our simulations the same orbital configurations, the G13 orbit.
This orbits geometry corresponds to an inclination of $i_1 = -109$ and a pericenter argument of $\omega_1 = 60$ for the first progenitor galaxy, i.e., in case of an unequal mass merger the more massive galaxy, and $i_2 = 180$ and $\omega_2 = 0$ for the second progenitor galaxy.
Both galaxies approach each other on parabolic orbits, with the merger taking place at about 1.5~Gyr after we started the simulation.
The simulations were evolved to 3~Gyr.
The G09 and G01 orbits were included to understand the impact of the orbit on the remnant.
For the G09 orbit, the parameters are $i_1 = -109$ and $\omega_1 = 0$ for the first progenitor galaxy and $i_2 = 180$ and $\omega_2 = 0$ for the second progenitor galaxy, the G01 orbit matches a geometry of $i_1 = 0$, $\omega_1 = 0$ and $i_2 = 180$, $\omega_2 = 0$.
All three orbits are parabolic, with a pericentric distance of $r_{\rm peri} = r_{\rm d,1} + r_{\rm d,2}$, with $r_{\rm d,1}$ and $r_{\rm d,2}$ the disk scale radii for the first respective the second progenitor galaxy. 
G13 and G09 are orbits where the orientations of the progenitor disks are not in the orbital plane, while G01 is a retrograde orbit with the progenitors being in the orbital plane.
For more details on these orbits see \citet{naab:2003ApJ...597..893N}.
Table~\ref{tab:binary} contains a summary of all simulation parameters used for this study.

\begin{table*}[p]
\caption{Binary merger simulation sample at a timestep of 3~Gyr}             
\label{tab:binary}      
\centering          
\begin{tabular}{c c c c c c c c c c c}
\hline\hline       
                      
Model & Ratio\tablenotemark{(a)} & Orbit\tablenotemark{(b)} & $f_{\mathrm{gas}}$\tablenotemark{(c)} & $M_{\mathrm{Gal}}$\tablenotemark{(d)} &  $M_{\mathrm{DM}}$\tablenotemark{(e)} & $N_{\mathrm{Gal}}$\tablenotemark{(f)} & $f_*^{\mathrm{new}}$\tablenotemark{(g)} & $R_{1/2}$\tablenotemark{(h)} & $f_{DM}$\tablenotemark{(i)} & $f_{DM}^{0.5}$\tablenotemark{(k)}\\ 

\hline
 11 OBH2 13    & 1:1 &  G13 & 0.2 &  1.28 & 2.84  & 712 870 & 10.71 & 4.67 & 0.24 & 0.14 \\ 
 11 OBH2 09    & 1:1 &  G09 & 0.2 &  1.29 & 2.82  & 713 997 & 11.21 & 4.45 & 0.23 & 0.13 \\ 
 11 OBH0 13    & 1:1 &  G13 & 0.0 &  1.33 & 2.71  & 742 490 &  0.00 & 6.14 & 0.29 & 0.18 \\
 11 OBHNB0 13  & 1:1 &  G13 & 0.0 &  0.97 & 2.78  & 548 320 &  0.00 & 7.56 & 0.46 & 0.32 \\ 

 mix 11 OBH2 13 & 1:1 & G13 & 0.2 &  1.45 & 2.97  & 809 734 & 11.79 & 5.34 & 0.26 & 0.15 \\

 31 OBH2 13     & 3:1 & G13 & 0.2 &  0.84 & 1.99  & 466 326 &  9.34 & 4.75 & 0.29 & 0.17 \\ 
 31 OBH8 13     & 3:1 & G13 & 0.8 &  0.76 & 2.06  & 416 201 & 50.84 & 2.40 & 0.15 & 0.08 \\
 31 ASF2 13     & 3:1 & G13 & 0.2 &  0.85 & 2.00  & 469 070 &  9.77 & 4.53 & 0.28 & 0.16 \\ 
 31 ASF2 01     & 3:1 & G01 & 0.2 &  0.85 & 2.01  & 474 622 & 11.63 & 3.96 & 0.24 & 0.13 \\ 
 31 OBH2 09 320 & 3:1 & G09 & 0.2 &  6.72 & 15.90 & 465 699 & 10.14 & 9.37 & 0.28 & 0.17 \\ 

\hline                  
\end{tabular}
\tablecomments{(a) initial mass ratio of the two galaxies;
(b) Orbit type according to \citet{naab:2003ApJ...597..893N};
(c) Initial gas fraction of the disks of the progenitor galaxies;
(d) Stellar masses within 10\% of the dark matter virial radius in $10^{11}M_{\odot}$;
(e) Dark matter masses within 10\% of the dark matter virial radius in $10^{11}M_{\odot}$;
(f) Number of stellar particles within 10\% of the dark matter virial radius;
(g) Fraction of newly formed stars since the start of the simulation in comparison to the total number of stars at the final output time $t=3\rm ~Gyr$;
(h) Effective radius of the stellar component of the galaxy, calculated as three dimensional half-mass radius, in kpc
(i) Fraction of dark matter relative to the stellar component within the half-mass radius
(k) Fraction of dark matter relative to the stellar component within 0.5$R_{1/2}$
}

\caption{CosmoZoom Ellipticals sample at $z=0$}
\label{tab:cosmoresims}
\centering
\begin{tabular}{c c c c c c c c c c}
\hline\hline

Model & $M_{\mathrm{*}}^{tot}$\tablenotemark{(a)} & $M_{\mathrm{Gal}}$\tablenotemark{(b)} &  $M_{\mathrm{DM}}$\tablenotemark{(c)} &  $N_{\mathrm{Gal}}$\tablenotemark{(d)} & $f_*^{\mathrm{new}}$\tablenotemark{(e)} & $R_{1/2}$\tablenotemark{(f)} & $f_{in situ}$\tablenotemark{(g)} & $f_{DM}$\tablenotemark{(h)} & $f_{DM}^{0.5}$\tablenotemark{(i)}\\

\hline
 0040\_2 & 25.98 & 5.00 & 19.02 &  84 786  & 2.02 & 12.91 & 23.07 & 0.43 & 0.29 \\
 0053\_2 & 16.26 & 6.95 & 19.25 & 117 809  & 1.43 & 13.03 &  ---  & 0.37 & 0.23 \\
 0069\_2 & 18.08 & 4.94 & 13.28 &  83 804  & 1.49 &  8.84 & 21.78 & 0.31 & 0.20 \\
 0089\_2 & 10.76 & 5.23 & 11.36 &  88 772  & 0.57 & 10.43 & 16.43 & 0.34 & 0.22 \\
 0094\_2 &  9.69 & 4.79 & 12.49 &  81 258  & 0.80 &  7.53 & 25.99 & 0.27 & 0.15 \\
 0125\_2 &  8.66 & 4.34 & 11.06 &  73 546  & 0.90 &  9.08 & 22.45 & 0.33 & 0.20 \\
 0162\_2 &  6.28 & 3.64 & 8.42  &  61 808  & 2.11 &  9.78 & 12.92 & 0.40 & 0.27 \\
 0163\_2 &  7.04 & 3.52 & 7.17  &  59 701  & 0.90 & 10.38 & 14.96 & 0.37 & 0.27 \\
 0175\_2 &  6.91 & 3.68 & 9.94  &  62 401  & 1.05 &  7.37 & 26.96 & 0.31 & 0.20 \\
 0190\_2 &  5.83 & 3.15 & 5.84  &  53 401  & 2.72 &  6.99 & 14.67 & 0.29 & 0.20 \\
 0204\_2 &  5.87 & 2.69 & 5.95  &  45 554  & 1.63 &  6.50 & 15.63 & 0.25 & 0.15 \\

 0204\_4 &  5.73 & 2.93 & 5.38  & 398 025  & 1.40 &  6.36 & 23.86 & 0.20 & 0.11 \\
 0215\_4 &  6.03 & 3.14 & 6.70  & 425 565  & 1.50 &  4.25 & 36.14 & 0.19 & 0.10 \\
 0408\_4 &  2.85 & 1.66 & 2.29  & 224 648  & 0.94 &  4.51 & 20.88 & 0.16 & 0.10 \\
 0501\_4 &  2.72 & 1.68 & 3.18  & 227 403  & 2.70 &  3.27 & 39.31 & 0.19 & 0.10 \\
 0616\_4 &  2.52 & 1.72 & 2.68  & 232 935  & 4.00 &  5.02 & 29.90 & 0.27 & 0.17 \\
 0664\_4 &  2.11 & 1.23 & 2.38  & 166 379  & 2.46 &  2.81 & 38.12 & 0.16 & 0.08 \\

\hline
\end{tabular}
\tablecomments{(a) Total stellar mass within the dark matter virial radius in $10^{11}M_{\odot}$;
(b) Stellar mass within 10\% of the dark matter virial radius in $10^{11}M_{\odot}$;
(c) Dark Matter mass within 10\% of the dark matter virial radius in $10^{11}M_{\odot}$;
(d) Number of stellar particles within 10\% of the dark matter virial radius;
(e) Fraction of newly formed stars since a redshift of 0.27, which is approximately 3~Gyr, for comparison with the major merger sample;
(f) Effective radius of the stellar component of the galaxy, calculated as three dimensional half-mass radius, in kpc
(g) Fraction of stars formed in situ taken from \citet{oser:2012ApJ...744...63O} in \%.
(h) Fraction of dark matter relative to the stellar component within the half-mass radius
(i) Fraction of dark matter relative to the stellar component within 0.5$R_{1/2}$
}

\caption{CosmoZoom companion sample at $z=0$}
\label{tab:cosmocomps}
\centering
\begin{tabular}{c c c c c c c c c}
\hline\hline

Model & $M_{\mathrm{Gal}}$\tablenotemark{(a)}& $M_{\mathrm{DM}}$\tablenotemark{(b)} & $N_{\mathrm{Gal}}$\tablenotemark{(c)} & $f_*^{\mathrm{new}}$\tablenotemark{(d)} & $R_{1/2}$\tablenotemark{(e)} & $v_{\mathrm{max}}$\tablenotemark{(f)} & $f_{DM}$\tablenotemark{(g)} & $f_{DM}^{0.5}$\tablenotemark{(h)}\\
\hline
 0040\_s1 &  3.28 & 10.13 & 55 581 & 2.05 & 6.62 & 445.0 & 0.28 & 0.18\\
 0069\_s1 &  1.32 &  3.57 & 22 451 & 1.15 & 4.31 & 317.5 & 0.30 & 0.20\\
 0069\_s2 &  1.67 &  5.34 & 28 393 & 1.52 & 4.54 & 339.5 & 0.30 & 0.20\\
 0069\_s3 &  1.42 &  3.71 & 24 013 & 3.72 & 3.84 & 340.0 & 0.29 & 0.19\\

\hline
\end{tabular}
\tablecomments{(a) Stellar mass within 10\% of the dark matter virial radius in $10^{11}M_{\odot}$;
(b) Dark Matter mass within 10\% of the dark matter virial radius in $10^{11}M_{\odot}$;
(c) Number of stellar particles within 10\% of the dark matter virial radius;
(d) Fraction of newly formed stars since a redshift of 0.27, which is approximately 3~Gyr, for comparison with the major merger sample;
(e) Effective radius of the stellar component of the galaxy, calculated as three dimensional half-mass radius, in kpc;
(f) Maximum circular velocity at infall into the parent halo
(g) Fraction of dark matter relative to the stellar component within the half-mass radius
(h) Fraction of dark matter relative to the stellar component within 0.5$R_{1/2}$
}

\end{table*}

\begin{table*}
\caption{Magneticum BCGs at $z=0$}
\label{tab:hydrocosmo}
\centering
\begin{tabular}{c c c c c c c c}
\hline\hline

Model & $M_{\mathrm{*}}^{tot}$\tablenotemark{(a)}& $M_{\mathrm{Gal}}$\tablenotemark{(b)} &  $M_{\mathrm{DM}}$\tablenotemark{(c)} & $N_{\mathrm{Gal}}$\tablenotemark{(d)} & $R_{1/2}$\tablenotemark{(e)} & $f_{DM}$\tablenotemark{(f)} & $f_{DM}^{0.5}$\tablenotemark{(g)}\\

\hline
 00 & 174.4 &  52.2 & 508.6 & 142 140 & 62.68 & 0.71 & 0.52 \\
 02 & 178.6 &  69.4 & 439.8 & 163 955 & 40.47 & 0.51 & 0.26 \\
 03 & 152.8 &  88.3 & 359.4 & 182 988 & 30.98 & 0.30 & 0.14 \\

\hline
\end{tabular}
\medskip
\tablecomments{(a) Total stellar mass within the dark matter virial radius in $10^{11}M_{\odot}$;
(b) Stellar mass within 10\% of the dark matter virial radius in $10^{11}M_{\odot}$;
(c) Dark Matter mass within 10\% of the dark matter virial radius in $10^{11}M_{\odot}$;
(d) Number of stellar particles within 10\% of the dark matter virial radius;
(e) Effective radius of the stellar component of the galaxy, calculated as three dimensional half-mass radius, in kpc.
(f) Fraction of dark matter relative to the stellar component within the half-mass radius
(g) Fraction of dark matter relative to the stellar component within 0.5$R_{1/2}$
}

\end{table*}

The mass resolution for all binary merger simulations is $M_{\rm DM} = 2.25 \times 10^6 M_{\odot} h^{-1}$ for dark matter particles and $M_{\rm gas} = M_{\rm stars} = 1.30 \times 10^5 M_{\odot} h^{-1}$ for gas and star particles.
The gravitational softening length was set to $\epsilon_{\rm DM} = 0.083 h^{-1} \rm \,kpc$ for dark matter particles and to $\epsilon_{\rm gas} = \epsilon_{\rm stars} = 0.02 h^{-1} \rm \,kpc$ for gas and star particles.
For all simulations we used $h = 0.71$ and a baryonic mass fraction of $\Omega_{\rm B} = 0.044$.

\subsection{Cosmological Zoom-In Simulations}
To study the formation of elliptical galaxies within the cosmological context, i.e., from multiple mergers, we analyzed a set of galaxies extracted from cosmological simulations of structure formation. 
Our sample consists of 17 zoom-in re-simulations of individual halos, hosting a central spheroidal galaxy at $z=0$.
Additionally, we included four companion ellipticals, i.e., massive spheroids that are substructures within larger halos.
We will refer to the central spheroidal galaxies as CosmoZoom Ellipticals hereafter, and to the substructures as CosmoZoom Companions.

The parent cosmological box of $72^3 h^{-3} \rm \,Mpc^3$ was simulated using $512^3$ DM particles with a particle mass of $M_{\rm DM} = 2 \times 10^8 M_{\odot} h^{-1}$ and a comoving gravitational softening length of $2.52 h^{-1} \rm \,kpc$.
A WMAP3 \citep{spergel:2007ApJS..170..377S}, $\Lambda$CDM cosmology was adopted, with $\sigma_8 =0.77$, $\Omega_{\Lambda} = 0.74$, $\Omega_{\rm m} = 0.26$ and $h = 0.72$ and an initial slope for the power spectrum of $n_s = 0.95$.

From this dark matter only simulation, halos of different masses, ranging from $10^{11} M_\odot h^{-1}$ to $10^{13} M_\odot h^{-1}$, were selected at $z=0$.
All dark matter particles closer than $2\times r_{200}$ to the halo center at any snapshot are traced back in time.
We replace all dark matter particles identified that way by dark matter and gas particles at higher resolution with $\Omega_{\rm DM} = 0.216$ and $\Omega_{\rm B} = 0.044$. 
The details of the re-simulation method are described in \citet{oser:2010ApJ...725.2312O}. 

These simulations use cooling for a primordial gas composition and star-formation but do not include any black hole treatment.
The initial conditions were created using GRAFIC and LINGERS \citep{bertschinger:2001ApJS..137....1B}.
The simulations were evolved from $z \sim 43$ to $z=0$.

To achieve a proper resolution even for the smaller halos we performed the re-simulations at two different resolutions.
The most massive halos were re-simulated with twice the spatial resolution of the original dark matter only box.
In these re-simulations the particle masses are $M_{\rm DM} = 2.1 \times 10^7 M_{\odot}h^{-1}$ and $M_{\rm gas} = M_{\rm stars} = 4.2 \times 10^6 M_{\odot}h^{-1}$ with gravitational softening set to $\epsilon_{\rm DM} = 0.89 h^{-1} \rm \,kpc$ and $\epsilon_{\rm gas} = \epsilon_{\rm stars} = 0.4 h^{-1} \rm \,kpc$ for dark matter, gas and star particles, respectively.
To study the effects of the gas physics and the stellar component on the dark matter, the halos of this re-simulation level were also re-simulated with dark matter only at the same resolution.

For the less massive halos we used four times the spatial resolution of the original box, and particle masses of $M_{\rm DM} = 3.6 \times 10^6 M_{\odot}h^{-1}$ for the dark matter particles and $M_{\rm gas} = M_{\rm stars} = 7.4 \times 10^5 M_{\odot}h^{-1}$ for the gas and star particles with gravitational softening set to $\epsilon_{\rm DM} = 0.45 h^{-1} \rm \,kpc$ and $\epsilon_{\rm gas} = \epsilon_{\rm stars} = 0.2 h^{-1} \rm \,kpc$ respectively.
Table~\ref{tab:cosmoresims} contains all ellipticals extracted from the re-simulations used for this study, and Table~\ref{tab:cosmocomps} contains the companion halos, labeled with the name of their host central halo followed by an S and the number of the substructure.

\subsection{Magneticum Pathfinder Simulation}

We also extracted central galaxies of clusters from a hydrodynamical, cosmological simulation (Magneticum Pathfinder, {\it Box3/hr}, Dolag et al 2012, in preparation).
These galaxies were extracted from a  $128^3 h^{-3}\rm \,Mpc^3$ Box simulated using $2\times576^3$ particles and adapting a WMAP7 \citep{komatsu:2011ApJS..192...18K} $\Lambda$CDM cosmology with $\sigma_8 =0.809$, $h = 0.704$, $\Omega_{\Lambda} = 0.728$, $\Omega_{\rm M} = 0.272$ and $\Omega_{\rm B} = 0.0456$ and an initial slope for the power spectrum of $n_s = 0.963$.
Dark matter particles have a mass of $M_{\rm DM} = 6.9 \times 10^8 M_{\odot} h^{-1}$, gas particles have a mass of approximately $M_{\rm Gas} = 1.4 \times 10^8 M_{\odot} h^{-1}$, depending on their enrichment history, and stellar particles have about $M_{\rm Stars} = 3.5 \times 10^7 M_{\odot} h^{-1}$, depending on the state of the underlying stellar population.
For all components the gravitational softening length is set to $3.75 h^{-1} \rm \,kpc$. 
In this simulation, one gas particle can form up to four stellar particles.

This simulation follows the gas using a low-viscosity SPH scheme to properly track turbulence \citep{dolag:2005MNRAS.364..753D} and the cooling and star-formation description follows the pattern of metal production \citep{tornatore:2004MNRAS.349L..19T,tornatore:2007MNRAS.382.1050T}, including in a self-consistent way the dependence of the gas cooling on the local metallicity \citep{wiersma:2009MNRAS.393...99W}.
It also includes the feedback caused by galactic winds \citep{springel:2003MNRAS.339..289S} as well as feedback from black holes according to \citet{springel:2005MNRAS.361..776S}.
  
We selected the three halos among the most massive ones of which the central galaxy contained at least $10^5$ star particles and was not currently undergoing a merger.
We will refer to these massive central cluster galaxies as Magneticum BCGs hereafter.
Table~\ref{tab:hydrocosmo} summarizes the properties of all three ellipticals.

\section{Results}\label{sec:analysis}
We identify all stars within $0.1R_{\rm vir}$ to belong to the central elliptical, similarly to \citet{oser:2010ApJ...725.2312O}, independent of the simulation type.
The galaxy masses used throughout this work are calculated according to these definitions.

For the companion ellipticals from the cosmological simulations, where significant parts of the dark matter have been stripped already and thus the virial radius cannot be obtained from the dark matter mass of the subhalo, we use the maximum circular velocity $v_{\rm max}$ to calculate the virial mass and the virial radius the halo had before the infall into the parent halo, using Equations~\ref{eq:virial1} and~\ref{eq:virial} with $v_{\rm max}$ as $v_{\rm vir}$. 
These companion ellipticals are particularly interesting since our observational comparison sample, the elliptical galaxies in the Coma cluster \citep{thomas:2007MNRAS.382..657T}, are embedded in the larger Coma cluster environment themselves and thus supposedly have a similar formation scenario to the companion ellipticals.

\begin{figure*}
\begin{center}
  \includegraphics[width=2\columnwidth]{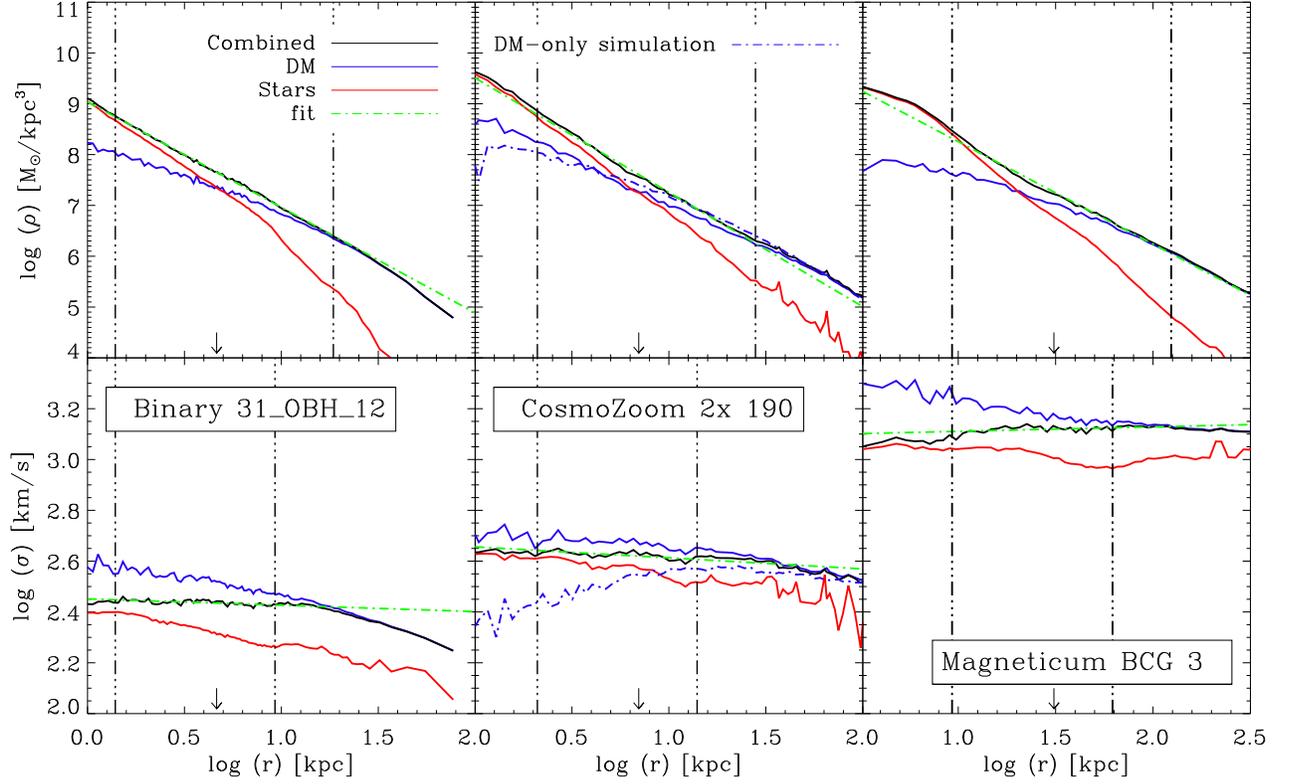}
  \caption{Density (upper panel) and velocity dispersion (lower panel) profiles for three example halos. Left panels: 3:1 Binary Elliptical $31\_\rm OBH2\_13$. Central panel: CosmoZoom Elliptical 190. Right panel: Magneticum BCG 03.
Red solid line: stellar profile; blue solid line: dark matter profile; black line: combined profile; green line: power-law fit to the combined profile. Dashed black lines for the left and central panel: $0.3 R_{1/2}$ and $4 R_{1/2}$ for the density, $0.3 R_{1/2}$ and $2 R_{1/2}$ for the velocity; for the right panel:$0.5 R_{1/2}$ and $4 R_{1/2}$ for the density, $0.5 R_{1/2}$ and $2 R_{1/2}$ for the velocity. The arrows mark the position of the half-mass radius. 
The dash-dotted blue line in the central panel shows the dark matter profile for the dark matter-only re-simulation of CosmoZoom Elliptical 190.}
  {\label{fig:denfit}}
\end{center}
\end{figure*}

For most of this work we use the elliptical galaxies at the final output time of 3~Gyr for the binary merger simulations and the ellipticals identified at a redshift of $z=0$ for all cosmological simulations.
First we calculate the three dimensional stellar half-mass radius $R_{1/2}$ for all elliptical galaxies in our sample, i.e., the radius which contains $50\%$ of all stars that were identified to belong to the galaxy. 
Next the intrinsic velocity dispersion and density within spherical equal-mass shells around the center of our simulated galaxies are calculated for stars and dark matter separately as well as for both components combined:
within each shell we first calculate the intrinsic velocity dispersion for stars and dark matter separately, as
\begin{flalign}
\sigma = \sqrt{ \sigma_r^2 + \sigma_\theta^2 + \sigma_\phi^2 }
\end{flalign}
with $\sigma_r = \sqrt{\langle v_r^2\rangle - \langle v_r\rangle ^2 }$ and $\sigma_\theta$ and $\sigma_\phi$ analogous.
Since dark matter and stellar particles have different masses within all our simulations, we need to calculate the total intrinsic mass-weighted velocity dispersion for the combined profiles within each shell as
\begin{flalign}
\sigma_{\rm tot} = \sqrt{\frac{m_{\rm DM}\sigma_{\rm DM}^2+m_{\rm S}\sigma_{\rm S}^2}{m_{\rm DM}+m_{\rm S}}}
\end{flalign}
to obtain a total velocity dispersion that is representative for the whole potential.
Profiles for three example halos are shown in Figure~\ref{fig:denfit}, for the density in the upper panels and the velocity dispersion in the lower panels.
We choose one example halo for each simulation type, for the Binary (left), the CosmoZoom (central) and the Magneticum (right) Ellipticals. 

A power law is fit to these velocity dispersion and density profiles, to stars and dark matter separately as well as to the combined profiles.
The fit to the combined profiles is shown as a green dashed line in Figure~\ref{fig:denfit} for the three example halos.
We were especially interested in the transition area, where the dark matter part becomes dominant in its contribution to the total density and velocity profiles.
This transition area is characterized well by the stellar half-mass radius $R_{1/2}$ for all galaxies, and thus we choose the fitting range to depend on $R_{1/2}$.
For the Binary Ellipticals and the CosmoZoom Ellipticals the best-fitting power-law index for the velocity dispersions is determined in the radius regime $0.3 R_{1/2}$ to $2 R_{1/2}$, the power law for the density is fit between $0.3 R_{1/2}$ and $4 R_{1/2}$.
For the CosmoZoom Companions and the Magneticum BCGs the best-fitting power laws are determined with a lower limit of $0.5R_{1/2}$ instead of $0.3R_{1/2}$ due to resolution limits.
The lower limits are chosen to ensure that the innermost particles included are at least 3 times the smoothing length away from the center.
The upper limits are chosen due to comparability with observations of Coma ellipticals by \citet{thomas:2007MNRAS.382..657T}, where density slopes are available up to $4 R_{\rm eff}$ while velocity dispersion slopes are only available up to $2 R_{\rm eff}$.

\begin{figure*}
	\begin{center}
  \includegraphics[width=2\columnwidth]{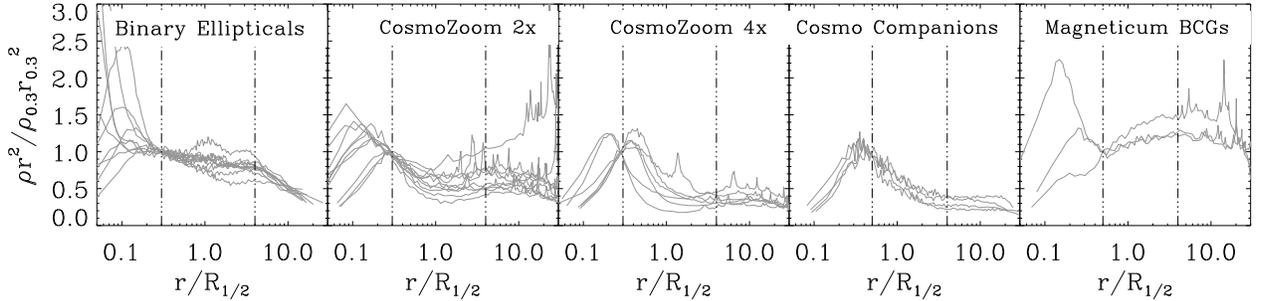}
  \caption{Combined dark matter and baryonic density profiles for all ellipticals, multiplied by $r^2$ and normalized to the density value at $0.3R_{1/2}$ ($0.5R_{1/2}$ for the CosmoZoom Companions and the Magneticum BCGs). From left to right: Binary Ellipticals, CosmoZoom Ellipticals with twice the spatial resolution, CosmoZoom Ellipticals with four times the spatial resolution, CosmoZoom Companions, and Magneticum BCGs. The dashed lines mark $0.3 R_{1/2}$ and $4 R_{1/2}$ for the first three panels, $0.5 R_{1/2}$ and $4 R_{1/2}$ in the last two panels.
}
  {\label{fig:denfit2}}
\end{center}
\end{figure*}

Figure~\ref{fig:denfit2} shows the combined dark matter and baryonic density profiles multiplied by $r^2$ for all Binary and CosmoZoom Ellipticals normalized to the density at $0.3R_{1/2}$ and for all CosmoZoom Companions and Magneticum BCGs normalized at $0.5R_{1/2}$, illustrating that a power-law fit is actually a good approximation for most of our ellipticals.
The only exceptions are the halos with an unusually dominant central stellar component, which are the least massive of the CosmoZoom Ellipticals with four times the spatial resolution and some of the CosmoZoom Companions. 
These halos have slightly curved combined density profiles and the best fitting power laws have generally steeper slopes. 

Figure~\ref{fig:denfit} and Figure~\ref{fig:denfit2} also show that the power-law fit to the total density profiles of the CosmoZoom Ellipticals and the Magneticum BCGs is a good approximation to even larger radii than $4 R_{1/2}$, even though the profiles become less smooth due to the presence of satellite galaxies.
This is in agreement with observations from weak lensing by \citet{gavazzi:2007ApJ...667..176G}, who found for 22 central elliptical galaxies that their total density profiles can be fit on average by a power law with a slope of $\gamma \approx -2$ over two decades in radius.
A similar behavior can also be seen for the CosmoZoom Companions, although their density profiles break down at about $10R_{1/2}$ due to the fact that their dark matter halos got stripped during the infall into the larger parent halo.
For the Binary ellipticals, the density profiles already break down at about $5R_{1/2}$ due to the fact that the simulations do not reach further out.

\subsection{Density and velocity dispersion slopes}
\begin{figure}
\begin{center}
  \includegraphics[width=\columnwidth]{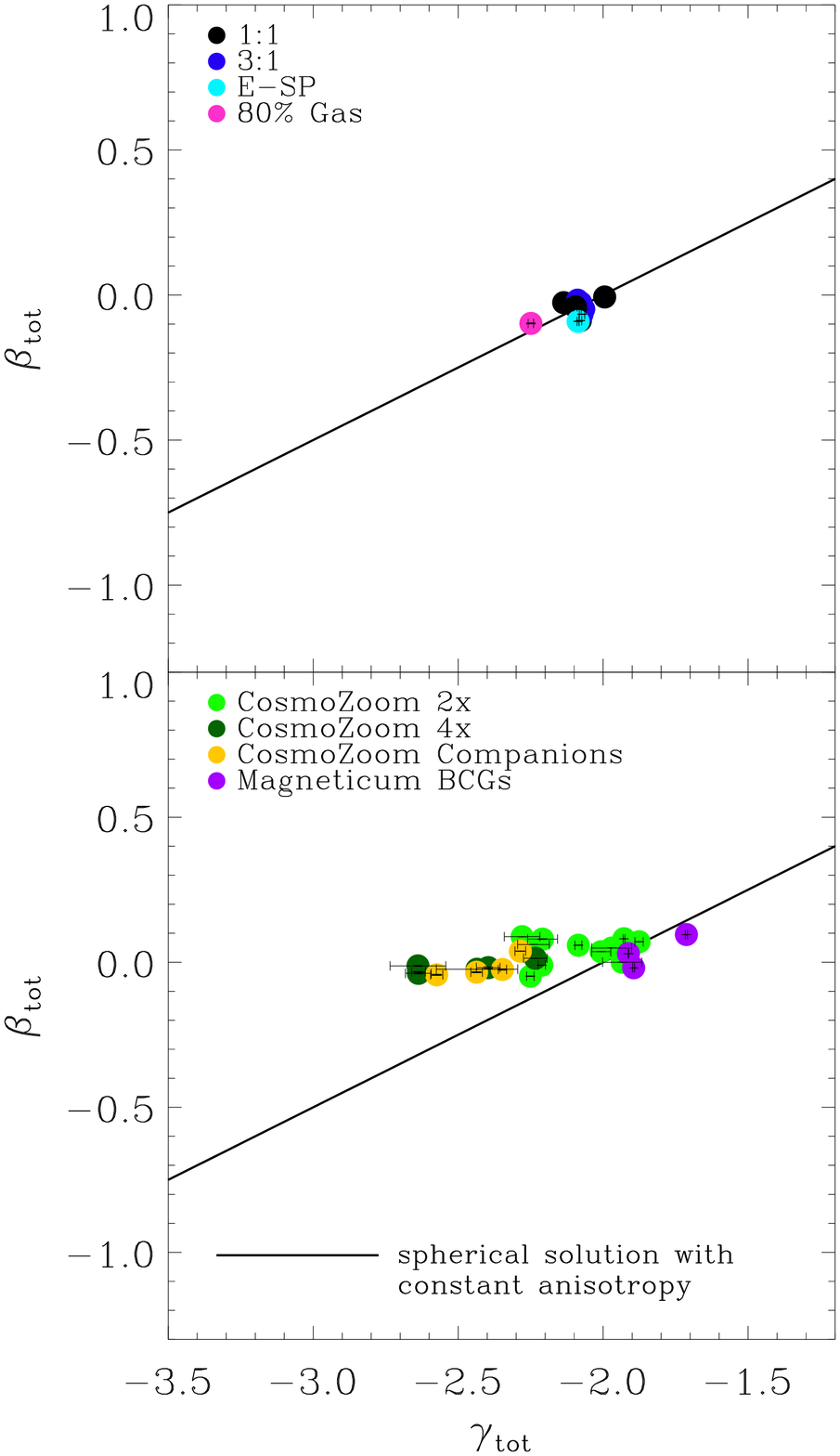}
  \caption{Slopes of the total velocity dispersions $\beta_{\rm tot}$ against the slopes of the total density profiles $\gamma_{\rm tot}$.
Upper panel: Results from the Binary Ellipticals: 1:1 spiral merger (black), 3:1 spiral merger (blue), Elliptical-spiral merger (cyan) and 3:1 spiral merger with 80\% gas (pink).
Lower panel: Results from the cosmological simulations: CosmoZoom 2X Ellipticals (bright green), CosmoZoom 4X Ellipticals (dark green), CosmoZoom Companions (yellow) and Magneticum BCGs (violet).
For all ellipticals the errors are RMS-deviations to the fit.
Black line: analytic solution for a spherical system with constant anisotropy.
}
  {\label{fig:densvel_aa}}
\end{center}
\end{figure}

\begin{figure}
\begin{center}
  \includegraphics[width=\columnwidth]{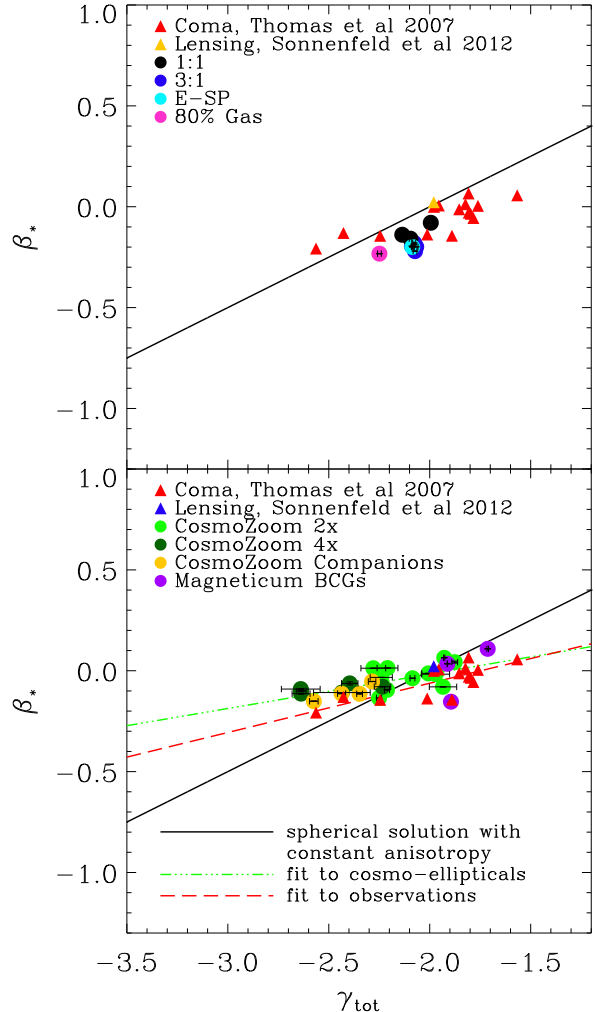}
  \caption{
Same as Figure~\ref{fig:densvel_aa} but for stellar velocity dispersion slopes versus total density slopes.
Red triangles show the total density and stellar velocity dispersion slopes for the Coma early-type galaxies presented in \citet{thomas:2007MNRAS.382..657T}.
The yellow (upper panel) respective blue (lower panel) triangle shows the slopes for the massive strong lensing early-type galaxy studied by \citet{sonnenfeld:2012ApJ...752..163S}.
Dashed lines: linear fits to the Coma values (red) and all cosmological simulations (green).
}
  {\label{fig:densvel_ast}}
\end{center}
\end{figure}

\begin{figure}
\begin{center}
  \includegraphics[width=\columnwidth]{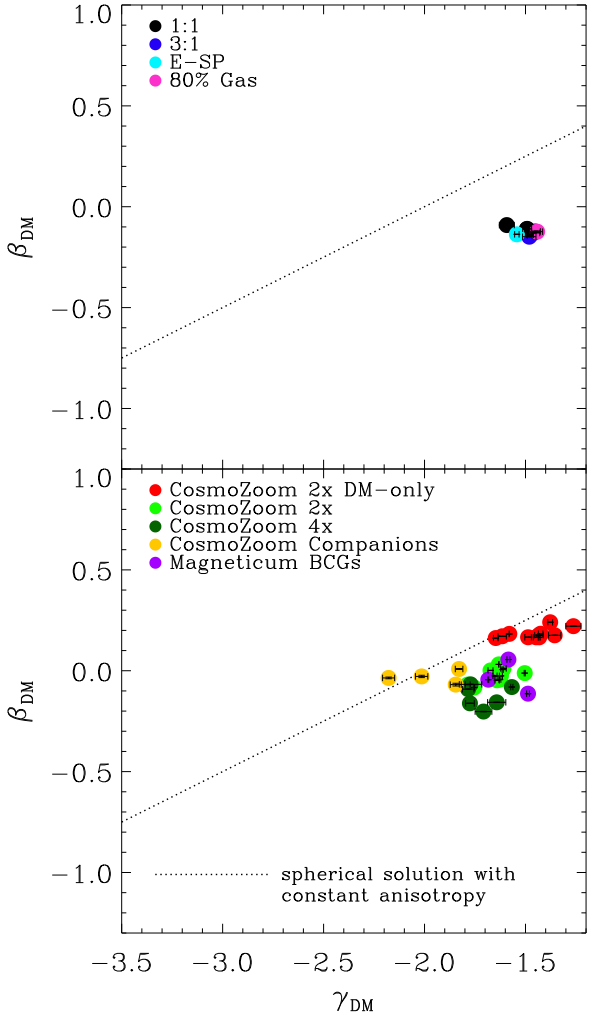}
  \caption{
Same as Figure~\ref{fig:densvel_aa} but for both velocity and density slopes of the dark matter alone.
Red circles: slopes of the velocity and density from the CosmoZoom dark matter-only re-simulations.
Dotted black lines: analytic solution for a spherical system with constant anisotropy.
The density slopes cluster in a range of $-1.8 \lesssim \gamma_{\rm DM} \lesssim -1.4$ excluding the CosmoZoom Companions (yellow circles) and the dark matter-only re-simulations (red), as explained in the text.
}
  {\label{fig:densvel_dd}}
\end{center}
\end{figure}

\begin{figure}
\begin{center}
  \includegraphics[width=\columnwidth]{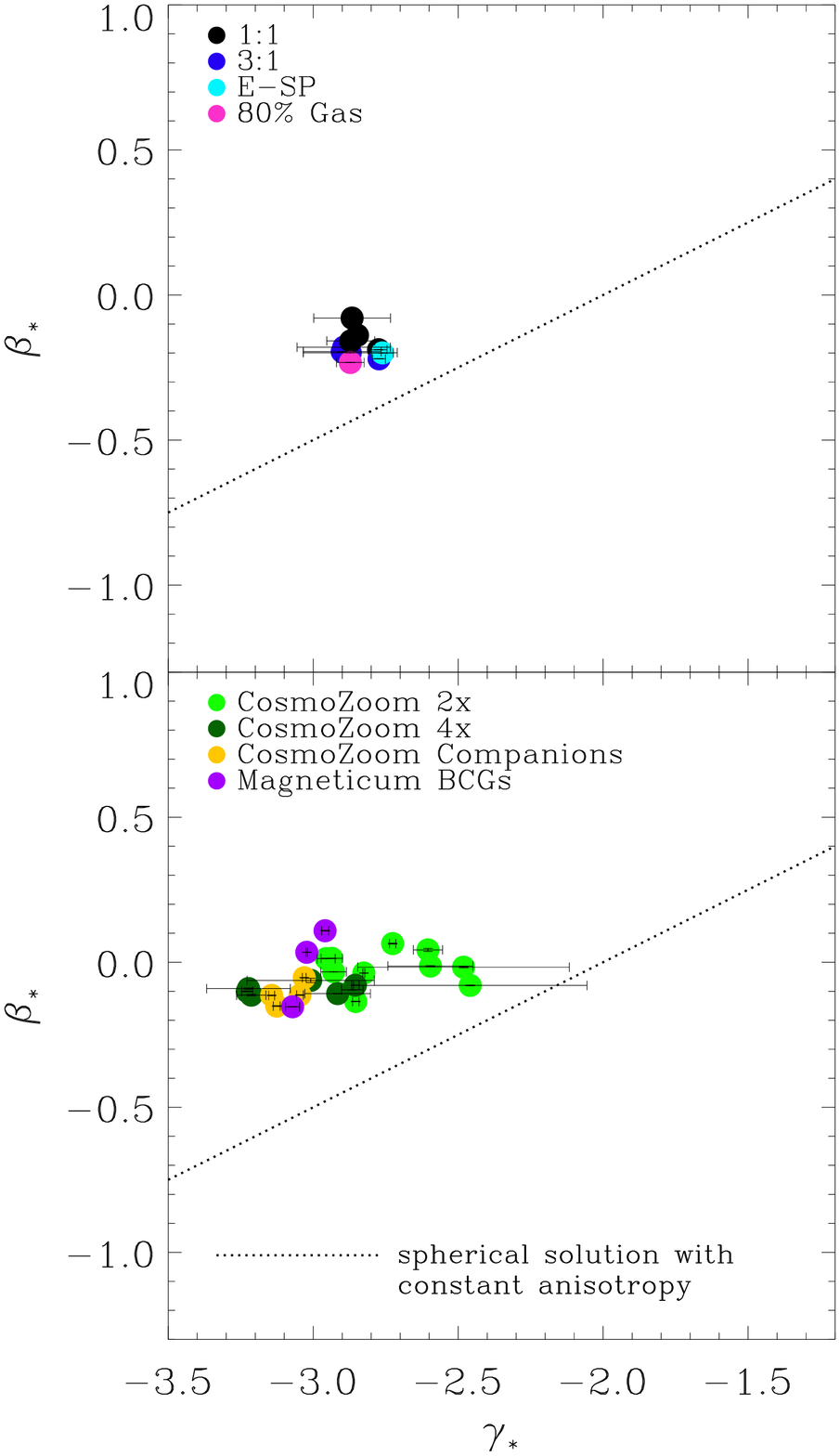}
  \caption{
Same as Figure~\ref{fig:densvel_aa} but for both velocity and density slopes of the stellar component alone.
Dotted black lines: analytic solution for a spherical system with constant anisotropy.
The slopes show no clustering in the density, they are spread over a range of $-3.25 \lesssim \gamma_{\rm *} \lesssim -2.4$.
}
 {\label{fig:densvel_stst}}
\end{center}
\end{figure}

For a spherical isothermal system the solution of the Jeans Equation is
\begin{flalign}
\rho (r) = \frac{\sigma ^2}{2\pi G r^2} .
\end{flalign}
More generally, under the assumption that both the velocity dispersion and the density can be described by simple power laws, i.e., $\rho(r) = Ar^{\gamma}$ and $\sigma(r) = Br^{\beta}$ with $A$ and $B$ being constants, the Jeans Equation has the following solution:
\begin{flalign}
\rho (r) = \frac{C \sigma (r)^2}{4\pi G r^2}
\end{flalign}
with $C$ being a dimensionless constant.
Thus, the slopes of the dispersion and the density are correlated by
\begin{flalign}\label{eq:slopes}
\beta = 0.5\gamma + 1 .
\end{flalign}
This correlation is shown in both panels of Figure~\ref{fig:densvel_aa} and Figure~\ref{fig:densvel_ast} as a solid black line and in Figure~\ref{fig:densvel_dd} and Figure~\ref{fig:densvel_stst} as dotted black line.
These relations hold for all systems with constant anisotropy as a function of radius, since different values for the constant anisotropy only change the constant $C$ and thus Equation~\ref{eq:slopes} does not change.


Figure~\ref{fig:densvel_aa} shows the total density slopes and total velocity dispersion slopes for all Binary Ellipticals in the upper panel and for all CosmoZoom Ellipticals and Magneticum BCGs in the lower panel. 
We can clearly see that all total slopes of the Binary Ellipticals lie close to the solution of the Jeans Equation, i.e., they have total density slopes around $\gamma_{\rm tot} = -2.1$ and total velocity dispersion slopes around $\beta_{\rm tot} = 0$, implying that these ellipticals are fairly close to spherical systems with constant anisotropy, or even isotropic in some cases.

For our sample, neither the choice of different orbits (G01, G09 or G13) nor the merger type (i.e., if it is a 3:1, a 1:1 or an E-SP merger) or the presence of a Black Hole changes the resulting density slopes significantly, while the variation of the initial gas fraction causes the only significant shift in the density slope.
The simulation with $80\%$ initial gas fraction has a significantly steeper total density slope than the comparable simulation with $20\%$ initial gas fraction.
At present day, mergers with high gas fractions are unlikely, but at a higher redshift gas-rich major mergers are much more frequent.
Still, our $80\%$ gas merger has very large and extended progenitor gas disks, which is unrealistic for high-$z$ disks.

The CosmoZoom Ellipticals show a much larger variety of total density slopes (see the lower panel of Figure~\ref{fig:densvel_aa}) than the Binary Ellipticals, although they also have flat total velocity dispersion curves, i.e., the slopes of the power-law fits are close to zero.
This is in agreement with a detailed analysis presented in \citet{lyskova:2012MNRAS.423.1813L}, using the same simulation set as \citet{oser:2010ApJ...725.2312O}.
While the CosmoZoom Ellipticals with the flatter density slopes around $\gamma_{\rm tot} = -1.9$ and total velocity dispersion slopes around $\beta_{\rm tot} = 0.05$ are close to the solution of the Jeans Equation and thus fairly close to spherical systems with constant anisotropy, the CosmoZoom Ellipticals with the steeper density profiles are not.
The steeper the total density slopes are, the larger is the deviation from the Jeans solution.
This could be due to a combination of gradients in the anisotropy and non-spherical effects that are not included in the simple spherically symmetric approach of this paper.
The details will be discussed in a subsequent paper (Remus et al, in preparation).
We also find that there is no difference between the behavior of the CosmoZoom Ellipticals and the CosmoZoom Companions. 
All three Magneticum BCGs are close to the Jeans solution and have relatively flat total density slopes compared to the majority of the CosmoZoom Ellipticals, with their total velocity dispersion slopes are flat as well.

Figure~\ref{fig:densvel_ast} shows the stellar velocity dispersion slopes $\beta_{\rm *}$ and the total density slopes $\gamma_{\rm tot}$ for the Binary Ellipticals in the upper and the CosmoZoom and Magneticum Ellipticals in the lower panel.
This figure also includes the stellar velocity dispersion and total density slopes obtained from observations of the Coma ellipticals as presented in \citet{thomas:2007MNRAS.382..657T} and the massive, strong-lensing early-type galaxy discussed in detail in \citet{sonnenfeld:2012ApJ...752..163S}.
The upper panel indicates that our limited sample of Binary Ellipticals cannot reproduce the range of slopes that is seen for the Coma cluster ellipticals, and the observations show no clustering around the values of the Binary Ellipticals.
Thus the scenario of a present-day major merger seems unlikely to be the dominant formation scenario for the Coma ellipticals.
Binary merger between two high-redshift spirals might produce different results, because the initial conditions for those galaxies would look very different, i.e., they would for example have no large stable gas disks and different dark matter halos, thus we cannot exclude high-redshift binary mergers to be the dominant formation mechanism for the Coma ellipticals.

As can be seen in the lower panel of Figure~\ref{fig:densvel_ast}, the range of values of the slopes of the CosmoZoom Ellipticals is similar to the range found for the observed Coma cluster ellipticals \citep{thomas:2007MNRAS.382..657T}, although there is a slight offset in the velocity dispersion slopes with respect to the observations.
The Magneticum BCGs are in good agreement with the slopes found for the more massive Coma Ellipticals, although the observational sample does not include the BCGs since the data of the Coma BCGs reach out to only $0.5R_{1/2}$.

Interestingly, if we look at the density slopes against the velocity slopes for the dark matter component only, the values cluster around $\gamma_{\rm DM} = -1.5$ for the Binary Ellipticals and around $\gamma_{\rm DM} = -1.67$ for the CosmoZoom Ellipticals and the Magneticum BCGs, as shown in Figure~\ref{fig:densvel_dd}.
This is in agreement with the observational results presented by \citet{sonnenfeld:2012ApJ...752..163S}, who found a dark matter density slope of $\gamma_{\rm DM} = -1.7 \pm 0.2$ for their massive strong-lensing early-type galaxy.
Only the CosmoZoom Companions show different values, which is most likely due to the fact that the outer parts of their dark matter halos have been stripped significantly during the infall in the parent halo, causing a steeper slope.

We also include in the lower panel of Figure~\ref{fig:densvel_dd} the slopes from the fits to the profiles of the dark matter only re-simulations of the CosmoZoom-2x Ellipticals, to study the influence of the stellar component on the dark matter halos directly.
As can be seen, the density slopes of the dark matter only re-simulations are slightly flatter (around $\gamma_\text{DM only} = -1.46$) than for the dark halos that contain a stellar component, and both their density and velocity dispersion slopes are closer to the theoretical solution for an isothermal sphere.
This can also be seen in the central panels of Figure~\ref{fig:denfit}, which shows the density and velocity dispersion curves of the dark matter component for an example CosmoZoom Elliptical as solid blue line and the corresponding curves for the dark matter only simulation of the same halo as the blue dashed line.
These values for the slopes of dark matter halos from dark matter only simulations are in agreement with results for the slopes of the central density of dark matter halos from high resolution dark matter only simulations presented by \citet{moore:1999MNRAS.310.1147M}, who found slopes of $\gamma_\text{DM only} \approx -1.5$, which is the same as what would be expected for NFW-profiles \citep{navarro:1996ApJ...462..563N} in this radius range.

We see that the presence of the stellar component significantly alters the distribution of the dark matter.
If the baryonic component is included in the simulation, the dark matter halo is denser in the center and thus the dark matter density slope is steeper, although for both the simulation with and without baryons the density converges to the same values at large radii beyond approximately 5$R_{1/2}$.
This is in agreement with results presented for example by \citet{onorbe:2007MNRAS.376...39O} and \citet{johansson:2012ApJ...754..115J}, and is due to the well-known effect of adiabatic contraction, i.e., the dark matter particles are pulled inward due to the condensation of the gas in the center of the halo (e.g., \citealp{jesseit:2002ApJ...571L..89J,blumenthal:1986ApJ...301...27B,gnedin:2004ApJ...616...16G,gnedin:2011arXiv1108.5736G}, see however \citealp{dutton:2007ApJ...654...27D} and, regarding the effects of expansion due to sudden outflows driven by supernovae, \citealp{pontzen:2012MNRAS.421.3464P}).
An even stronger effect caused by the presence of baryons can be seen for the velocity dispersion profile in the lower panel of Figure~\ref{fig:denfit}.
While the velocity dispersion of the dark matter component of the simulation with baryons slightly decreases with larger radii, the dark matter-only simulation shows a velocity dispersion that is strongly increasing with larger radii up to approximately 2$R_{1/2}$ and thus the power-law fit has a positive slope.

Figure~\ref{fig:densvel_stst} shows the density and velocity dispersion slopes for the stellar component, with the Binary Ellipticals in the upper and the CosmoZoom and Magneticum BCGs in the lower panel.
For all ellipticals, the density slopes of the stellar component are generally steeper than the slopes of the dark matter component.
While the slopes of the stellar component for the Binary Ellipticals are all around $\gamma_{\rm *} = -2.9$, the slopes of the stellar component of the CosmoZoom Ellipticals and the Magneticum BCGs show a different behavior:
In contrast to the dark matter component, which has a small range of density slopes from $-1.8 \lesssim \gamma_{\rm DM} \lesssim -1.4$ excluding the Companion ellipticals, the stellar component covers a larger range of slopes ($-3.25 \lesssim \gamma_{\rm *} \lesssim -2.4$) and there is no correlation between the stellar slope and the type of simulation.
This implies that the stellar component is responsible for the steepness of the total density slope, namely a more dominant stellar component in the center of a galaxy leads to a steeper total density slope.

\subsection{The influence of gas and star formation on the density slope}
Unlike the stellar and dark matter particles, which are collisionless, the gas particles can dissipate their energy and thus condense in the center of the galaxy.
Thus, new stars are mostly formed in the central area of the galaxy.
The more gas that is present in a merger event, the more dominant is this newly formed stellar component in the center.

\begin{figure}
\begin{center}
  \includegraphics[width=\columnwidth]{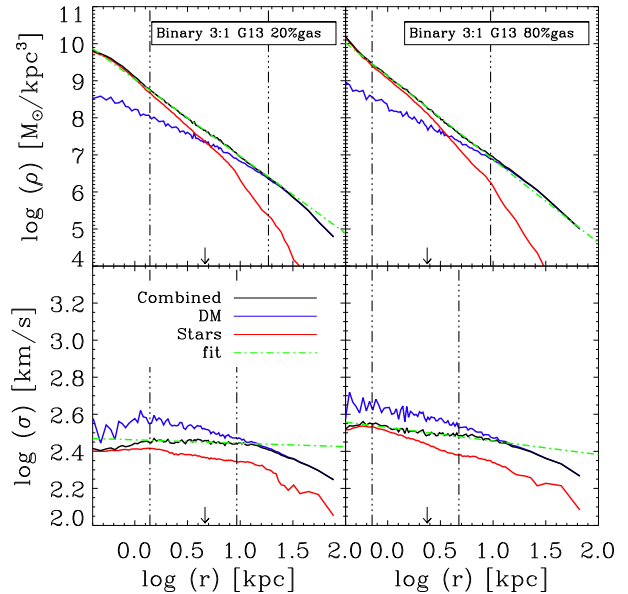}
  \caption{Same as Figure~\ref{fig:denfit} for the 3:1 Binary Elliptical with 20\% initial gas fraction (left panel) and the 3:1 Binary Elliptical with 80\% initial gas fraction (right panel).
}
  {\label{fig:denapp}}
\end{center}
\end{figure}

This effect is nicely demonstrated in the comparison between the two Binary Ellipticals which have identical initial conditions apart from the initial gas fraction of 20\% respective 80\%.
Figure~\ref{fig:denapp} shows the density and velocity dispersion profiles for both ellipticals.
In case of the merger with 80\% gas fraction we see two effects that change the stellar density profile compared to the case of the merger with a gas fraction of 20\%:
First, we see that the overall density is generally higher since more stars have been formed over the whole radius, i.e., the normalization $A$ in the profile $\rho(r) = Ar^{\gamma}$ is larger, while the dark matter profile for both halos did not change significantly.
Second, the central part of the elliptical is much more compact, as a large amount of stars has been formed there.
Both effects together cause the total density profile to be much steeper in case of the 80\% gas merger.
The half-mass radius of the 80\% gas merger is with $R_{1/2} = 2.4\rm ~kpc$ just half as large as the half-mass radius of the 20\% gas merger, while the fraction of stars that are formed during the merger is with $f_*^{\rm new} = 51\%$ much higher than in the 20\% gas merger ($f_*^{\rm new} = 9\%$), as can be seen in Table~\ref{tab:binary}.

\begin{figure*}
\begin{center}
  \includegraphics[width=2\columnwidth]{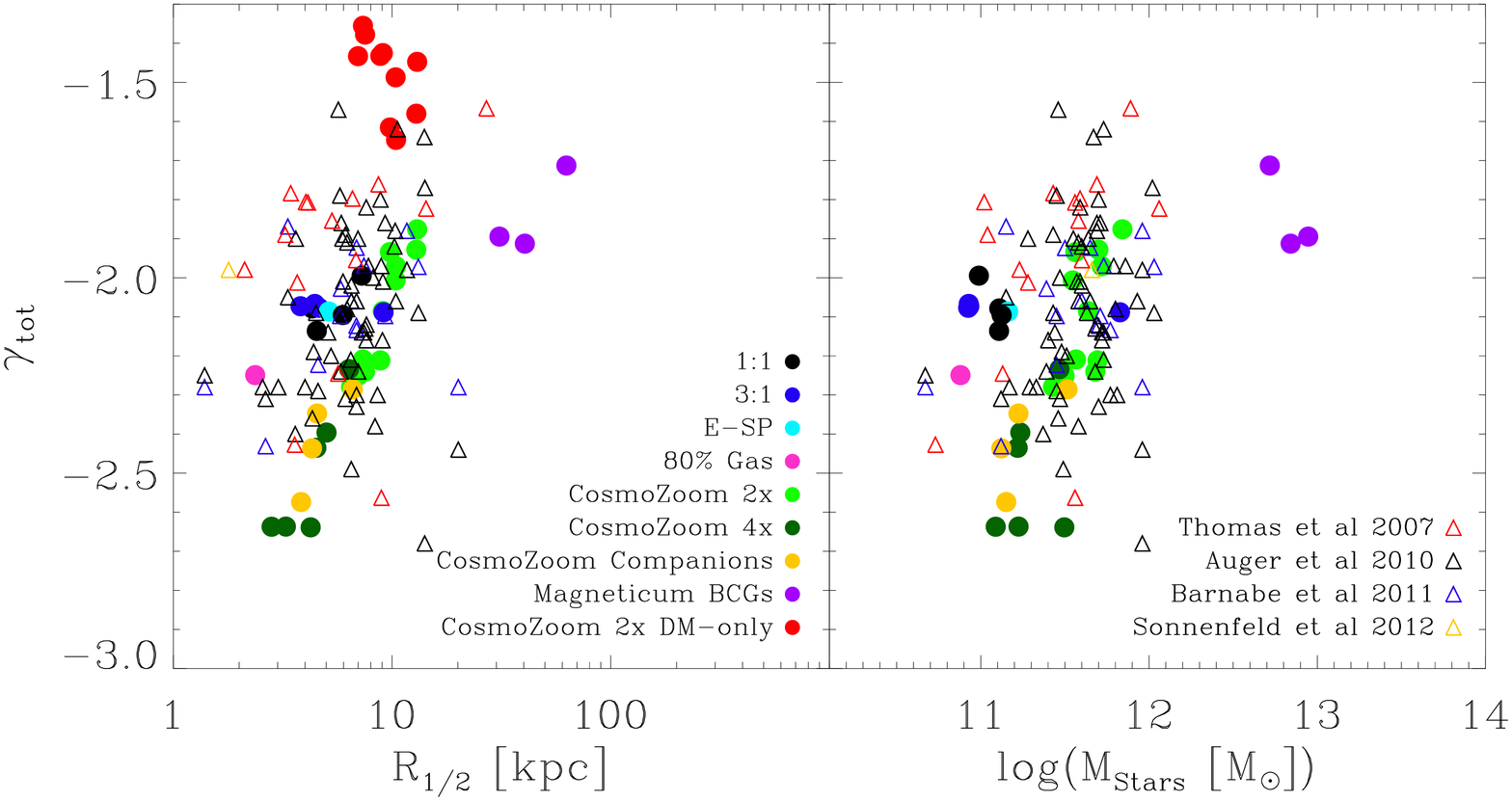}
  \caption{Left: total density slope against the half-mass radius for all our ellipticals.
Right: total density slope against the stellar mass within $R_{1/2}$.
Colors are the same as in Figure~\ref{fig:densvel_dd}.
Red triangles are observations of Coma ellipticals from \citet{thomas:2007MNRAS.382..657T}, blue open triangles are lensing results from \citep{barnabe:2011MNRAS.415.2215B}, black open triangles are lensing results from \citep{auger:2010ApJ...724..511A} and the yellow open triangle represents the results from \citet{sonnenfeld:2012ApJ...752..163S}.
We use the effective radius as half-mass radius for the observed galaxies.
}
  {\label{fig:denslop}}
\end{center}
\end{figure*}
\begin{figure}
\begin{center}
  \includegraphics[width=\columnwidth]{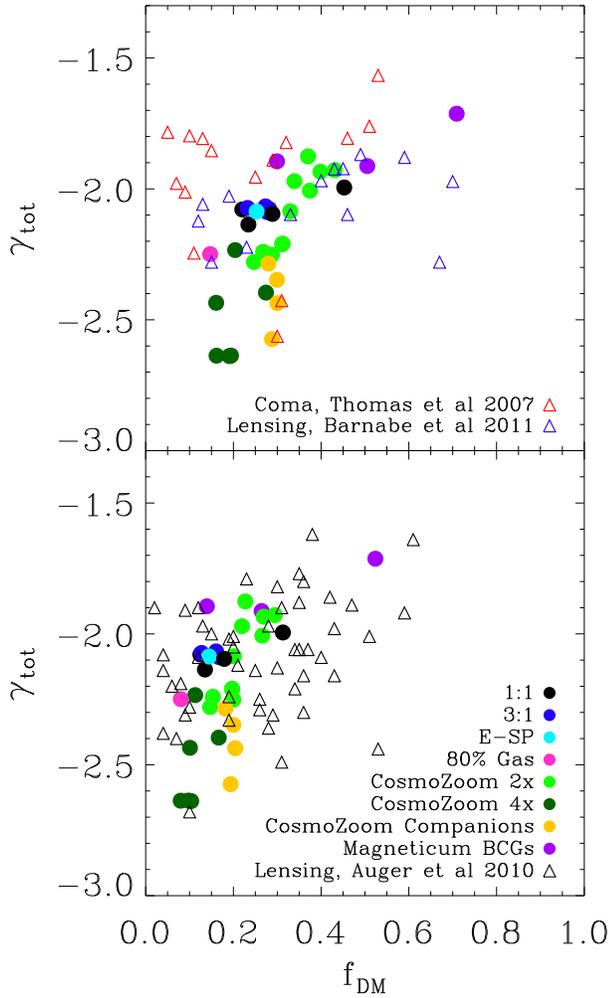}
  \caption{Total density slope against the fraction of dark matter within $R_{1/2}$ (upper panel) or within $0.5R_{1/2}$ (lower panel).
The red triangles are observations of Coma ellipticals from \citep{thomas:2007MNRAS.382..657T}, the blue open triangles are lensing results from \citep{barnabe:2011MNRAS.415.2215B}, the black open triangles are lensing results from \citep{auger:2010ApJ...724..511A}.
}
  {\label{fig:fdm}}
\end{center}
\end{figure}

This correlation between the steepness of the total density slope and the half-mass radius can also be seen for the CosmoZoom Ellipticals and Companions, as shown in the left panel of Figure~\ref{fig:denslop}.
The steeper the total density slope of an elliptical, the smaller the half-mass radius.  
We also find a (weaker) correlation between the stellar mass of an elliptical galaxy and its total density slope $\gamma_{\rm tot}$, as shown in the right panel of Figure~\ref{fig:denslop}, and a correlation between the total density slope and the dark matter fraction within the half-mass radius, as shown in Figure~\ref{fig:fdm}.
As expected, the total density slope is steeper the more dominant the stellar component is compared to the dark matter in the inner part of the galaxy, i.e., the smaller the fraction of dark matter within the half-mass radius.

In these figures we included the results for the Coma ellipticals by \citet{thomas:2007MNRAS.382..657T} as well as the results from the SLACS strong lensing survey presented by \citet{auger:2010ApJ...724..511A}, \citet{barnabe:2011MNRAS.415.2215B} and \citet{sonnenfeld:2012ApJ...752..163S}, and we see that our result are in good agreement with the observations.
The only exceptions are the Magneticum BCGs for which we have no observational counterparts, neither in mass nor in half-mass radius.

Also shown in the left panel of Figure~\ref{fig:denslop} are the total density slopes against the half-mass radii for the CosmoZoom dark matter-only simulations.
As seen before, the dark matter-only simulations show much flatter density slopes than the simulations including baryon physics, i.e., adding the baryons steepens the total density profile of the halos. We find that the observations clearly favor the slopes given by simulations with baryon physics.

There are a few early-type galaxies in the Coma observational sample that have a slope around $\gamma_{\rm tot} \approx -2$ and a very low dark matter fraction (see the upper panel of Figure~\ref{fig:fdm}).
This kind of ellipticals with very low dark matter fractions are also present in the strong lensing sample of \citet{auger:2010ApJ...724..511A} (see the lower panel of Figure~\ref{fig:fdm}), but not in the strong lensing sample of \citet{barnabe:2011MNRAS.415.2215B}.
We cannot reproduce these early-type galaxies with any of our simulated scenarios, not even with the CosmoZoom Companions.
Those early types seem to have a dominant stellar component, but a relatively flat density slope.

In case of the Coma Cluster ellipticals we know from \citet{thomas:2011MNRAS.415..545T} that these ellipticals with low dark matter fractions have large dynamical mass-to-light ratios in the case of the Coma Cluster ellipticals compared to a Kroupa IMF.
This means that these ellipticals either have a bottom-heavy stellar initial mass function or that their dark matter density is nearly identical to the density of the stellar component.

There have been several recent papers indicating especially in case of massive early-type galaxies, that the IMF is not universal but variable (\citealp{cappellari:2012Natur.484..485C,ferreras:2013MNRAS.429L..15F,vanDokkum:2011ApJ...735L..13V,conroy:2012ApJ...760...71C,vanDokkum:2012ApJ...760...70V,treu:2010ApJ...709.1195T}).
This is interesting since the predicted dark matter fractions strongly depend on the assumed IMF. 
For example, in case of an IMF like Kroupa, the observed dark matter fractions for the Coma Cluster ellipticals would be much higher (between 40\% and 70\%, see \citealp{thomas:2011MNRAS.415..545T}) and thus fit quite well to our results from the simulations.
However, as shown by \citet{conroy:2012ApJ...760...71C} and \citet{wegner:2012AJ....144...78W}, not even a variable stellar IMF can always explain the high dynamical mass-to-light ratios that are observed.

On the other hand, if the dark matter density follows the stellar component closely enough, both components become indistinguishable from each other and thus the stellar mass becomes overestimated in the Schwarzschild modeling.
This is explained in detail for the early-type galaxies in the Abell 262 cluster in \citet{wegner:2012AJ....144...78W}.
One possible way to explain an increase of the dark matter is by adiabatic contraction, as discussed beforehand.
Nevertheless, the contraction would have to be very strong, stronger then what is seen in our simulations.
Another process that could cause similar stellar and dark matter densities is violent relaxation, which has been discussed in \citet{wegner:2012AJ....144...78W} as well.

It is also possible that our simulations simply do not include ellipticals that are the equivalents to those observed early types.
It is possible that these early-type galaxies are actually spirals that suffered from tidal and gas stripping while they entered a dense environment like, for example, the Coma cluster.
\citet{gunn:1972ApJ...176....1G} already suggested ram pressure stripping to be an efficient way to form S0 early-type galaxies in dense environments, and ongoing stripping has been recently observed for the Virgo Cluster by \citep{abramson:2011AJ....141..164A}, but we have no simulations of such an event in our sample.

\begin{figure}
\begin{center}
  \includegraphics[width=\columnwidth]{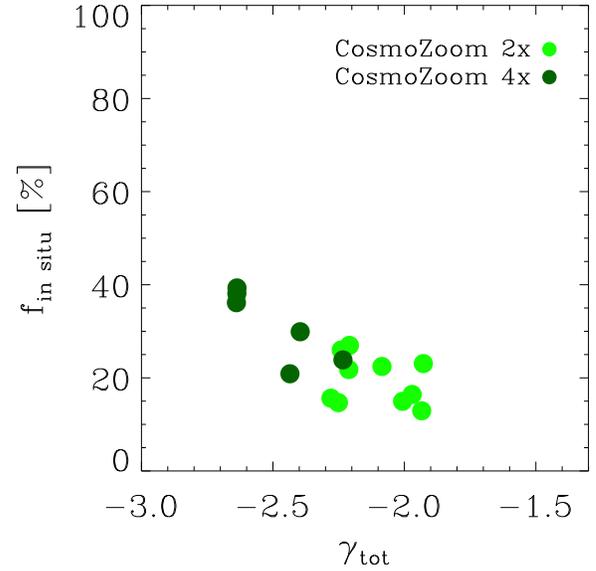}
  \caption{Fraction of stars formed in situ taken from \citet{oser:2012ApJ...744...63O} versus the total density slope of the halos for the ellipticals taken from the re-simulations with twice (four times) the spatial resolution in bright green (dark green) circles.
}
  {\label{fig:insitu}}
\end{center}
\end{figure}
For the CosmoZoom Ellipticals that have been studied in \citet{oser:2012ApJ...744...63O} we found that the slope of the total density correlates with the fraction of stars formed in situ, see Figure~\ref{fig:insitu}.
The more stars have been formed within the galaxy itself the steeper the slope of the total density, while the accretion of stars by merger events flattens the slope.
This is in agreement with the fact that for the Binary Ellipticals the steepest slope can be found for the $80\%$ gas merger, as discussed above.

\subsection{Evolution of the slopes}\label{sec:slopeevolution}
\begin{figure}
\begin{center}
  \includegraphics[width=\columnwidth]{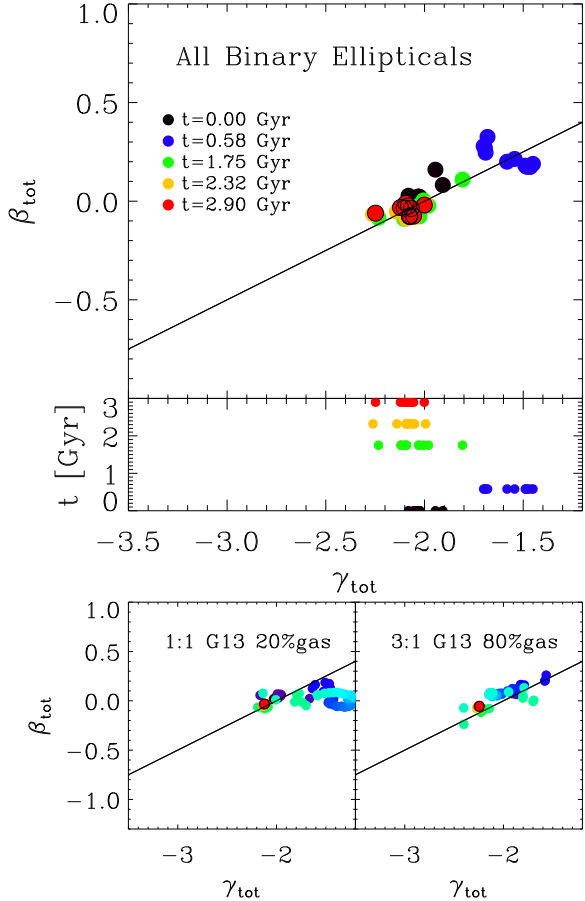}
  \caption{Total velocity slopes versus total density slopes for all binary merger ellipticals.
Upper panel: Slopes at different time-steps for all ellipticals at $t=0\rm ~Gyr$ (black circle), $t=0.58\rm ~Gyr$ (blue circle), $t=1.75\rm ~Gyr$ (green circle), $t=2.32\rm ~Gyr$ (orange circle) and $t=2.9\rm ~Gyr$ (red circle), with the merger event occurring around $1.5\rm ~Gyr$.
Central panel: Total velocity slopes versus time in Gyr. The colors are the same as in the upper panel.
Lower left panel: Isolated full evolution track from $t=0\rm ~Gyr$ to $t=3\rm ~Gyr$ for the 1:1 spiral merger on the G13 orbit with 20\% initial gas fraction.
Lower right panel: Isolated full evolution track from $t=0\rm ~Gyr$ to $t=3\rm ~Gyr$ for the 3:1 spiral merger on the G13 orbit with 80\% initial gas fraction.
The black lines in all figures show, like in Figure~\ref{fig:densvel_aa}, the analytic solution for a spherical system with constant anisotropy.}
  {\label{fig:evolution_binary}}
\end{center}
\end{figure}

To understand the origin of the total density slopes we study the time evolution of the total slopes.
In the upper panel of Figure~\ref{fig:evolution_binary} we show the evolution of the density and velocity dispersion slopes for all our Binary Ellipticals, at the time-steps $t=0\rm ~Gyr$ (initial condition time-step), $t=0.58\rm ~Gyr$, $t=1.75\rm ~Gyr$, $t=2.32\rm ~Gyr$ and $t=2.9\rm ~Gyr$.
Before the merger event occurs we fit the slopes to one of the spiral galaxies, in case of an unequal mass merger to the more massive spiral.
As soon as the galaxies are merged we fit the slopes to the remnant elliptical.
The first passage usually takes place between $t=0.5\rm ~Gyr$ and $t=0.6\rm ~Gyr$, the merger event around $t=1.5\rm ~Gyr$, thus only for the initial condition time $t=0\rm ~Gyr$ and the time-step $t=0.58\rm ~Gyr$ we actually fit the slopes to the spiral progenitor galaxies. 

\begin{figure}
\begin{center}
  \includegraphics[width=\columnwidth]{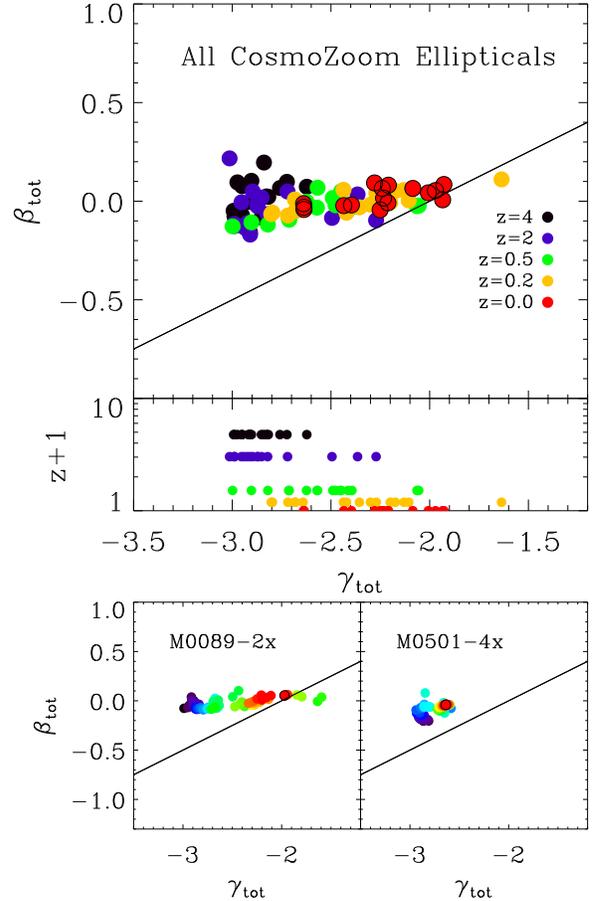}
  \caption{Total velocity slopes versus total density slopes for all ellipticals selected from the cosmological re-simulations.
Upper panel: Slopes at different redshifts for all ellipticals: $z=4$ (black circle), $z=2$ (blue circle), $z=0.5$ (green circle), $z=0.2$ (orange circle) and $z=0$ (red circle).
Central panel: Total velocity slopes versus redshift. The colors are the same as in the upper panel.
Lower left panel: Isolated evolution track for the halo M0089-2x, a massive elliptical which is growing through multiple mergers.
Lower right panel: Isolated evolution track for the halo M0501-4x, a low-mass elliptical with only minor mergers and smooth accretion.
The black lines in all figures show, like in Figure~\ref{fig:densvel_aa}, the analytic solution for a spherical system with constant anisotropy.}
  {\label{fig:evolution_cosmos}}
\end{center}
\end{figure}

Since we set the initial conditions for the Binary Ellipticals to fit present-day spirals, they are already isothermal systems at $t=0\rm ~Gyr$, i.e., the initial condition slopes are around $\gamma_{\rm tot} = -2.0$ and $\beta_{\rm tot} = 0$.
During the merger event, between $t=0.5\rm ~Gyr$ and $t=1.5\rm ~Gyr$, both the total density and velocity dispersion slopes are disturbed and not in equilibrium, with flatter slopes for the density and slightly positive slopes for the velocity dispersion.
This shows clearly that the flat density profiles during the merger are due to non-equilibrium effects, and that, as soon as the central parts of the spirals have merged and the central part of the elliptical has formed, both slopes return to the isothermal solution and stay (nearly) constant.

Detailed examples for two evolution tracks from $t=0\rm ~Gyr$ to $t=3\rm ~Gyr$ are shown in the lower two panels of Figure~\ref{fig:evolution_binary}.
The left panel shows the 1:1 spiral merger with 20\% initial gas fraction on a G13 orbit.
The right panel shows the 3:1 spiral merger with 80\% initial gas fraction on a G13 orbit, the one elliptical from the binary simulations that has a significantly steeper total density slope at $3\rm ~Gyr$ than all others.

The 1:1 spiral merger with 20\% initial gas fraction shows the typical behavior described above:
The merger event strongly disturbs the total density slope toward flatter slopes, while the velocity dispersion slope only changes slightly, and after the merger event takes place the slopes return to the initial configuration of an isothermal sphere.
The merger with an initial gas fraction of 80\% is the only merger event in our sample that behaves differently:
It evolves along the black line showing the isothermal solutions of the Jeans Equation, and the slope of the density never reaches equally flat values as the merger with 20\% initial gas fraction.
Even more importantly, after the merger event takes place the final elliptical does not have the same total density and velocity dispersion slopes as the initial setup.
Caused by a strong star formation due to gas condensation in the center of the newly formed elliptical, as discussed above, the final density slope is steeper than the density slope of the initial setup.
We conclude that a high gas fraction which causes a significant amount of star formation in the central part of the galaxy by condensing gas in the galaxies center, is needed for a merger event to cause a steeper total density slope once the progenitor system has reached an isothermal configuration.

Figure~\ref{fig:evolution_cosmos} shows in the upper panel the evolution of all the CosmoZoom Ellipticals, at redshifts $z=4$, $z=2$, $z=0.5$, $z=0.2$ and $z=0$.
We always fit the slopes of the most massive progenitor of the present-day elliptical at the given redshift.
The lower two panels show the full evolution tracks from $z=4$ to present day for two example ellipticals, one with a multiple merger history (left panel) and one with a smooth accretion history, where the merger only occur at high redshifts (right panel).

We see that the progenitors of the CosmoZoom Ellipticals have much steeper density profiles at high redshifts, evolving toward the isothermal spherical system case with every merging event, while the velocity dispersion slopes only change slightly during merger events and otherwise stay constant.
Between $z=4$ and $z=2$ the total density slopes do not change much, they are all about $\gamma_{\rm tot} \approx -3$, and they do not show an equally broad range of values, especially at $z=4$, than their present-day counterparts.
After $z=2$ the total density slopes of all progenitors become flatter and the range of slopes broadens.
This can be explained by the two-phase formation of galaxies that was introduced by \citet{oser:2010ApJ...725.2312O}: at redshifts $z\gtrsim 2$ the formation of the galaxies is dominated by the accretion of gas and the in situ star formation followed by a formation phase in which most of the mass growth of the galaxies is due to the accretion of stars in the form of satellite systems.

This also explains the correlation between the total density slopes and the fraction of stars formed in situ (see Figure~\ref{fig:insitu}) and the correlation between the total density slope and the half-mass radius $R_{1/2}$ (see Figure~\ref{fig:denslop}).
The CosmoZoom Ellipticals which have a steep density slope have a less dominant mass accretion by mergers in the second phase of their formation than those which have relatively flat total density slopes, i.e., their in situ fraction is higher and they are more compact.
CosmoZoom Elliptical M0501-4x in the lower right panel of Figure~\ref{fig:evolution_cosmos} is an example for such a merger poor second phase of formation.
Its total density slope hardly changes from $z=4$ to present day, and it has no really massive merger event in its formation history.

The opposite case is shown in the lower left panel of Figure~\ref{fig:evolution_cosmos}. 
CosmoZoom Elliptical M0089-2x has several minor merger events with mass ratios between 10:1 and 3:1 between $z=2$ and $z=0.8$ and even a major merger around $z=0.65$.
The effect of this major merger event can be seen in the lower left panel of Figure~\ref{fig:evolution_cosmos} as the green circles with the flattest total density slopes.
As for our Binary Ellipticals, the major merger event disturbs the total density slope even beyond $\gamma_{\rm tot} = -2$.
As for the Binary mergers, we also see for the CosmoZoom ellipticals that density slopes lower than $\gamma_{\rm tot} \approx -2$ are due to distortions and non-equilibria during massive merging events.

From both elliptical formation scenarios studied in this paper we conclude that merger events, other than in situ star formation, flatten the total density slopes toward the isothermal solution with a density slope of $\gamma_{\rm tot} \approx -2$.
Once the ellipticals have reached this configuration, they stay at this density distribution and, when disturbed, evolve back to a density distribution with a slope of $\gamma_{\rm tot} \approx -2$, even though the stellar and the dark matter component themselves might have changed.
We therefore conclude that total density distributions with a slope of $\gamma_{\rm tot} \approx -2$ act as an attractor solution.
Only a very gas rich merger event can steepen the slope again once it reached the attractor solution, as during a gas rich merger the gas can condense in the center of the elliptical and cause significant in situ star formation, but in the second phase of elliptical galaxy formation, which is dominated by the accretion and not by the formation of stars, a gas rich merger is not a common event.

\section{Summary and Discussion}\label{sec:conclusions}
We have investigated a set of 35 spheroidal galaxies formed from isolated binary merger events as well as in cosmological (zoom) simulations.
The isolated binary merger spheroids are taken from \citet{johansson:2009ApJ...707L.184J} and \citet{johansson:2009ApJ...690..802J}.
The spheroids formed in cosmological simulations are selected from two different simulations: The cosmological zoom-in simulation sample is a subset of the sample presented in \citet{oser:2010ApJ...725.2312O}, while the brightest cluster galaxy (BCG) sample is selected from a new full hydrodynamical cosmological simulation (Dolag et al., in preparation).
We analyzed the total intrinsic density and velocity dispersion profiles of the galaxies to investigate the dependence of the profiles on the formation history, mass, size and dark matter fraction.

We find that all our galaxies are close to isothermal, i.e their total (dark matter plus stellar) velocity dispersion distributions are flat, independent of the individual total, stellar or dark matter density distributions, the mass, the half-mass radius, the dark matter fraction within one half-mass radius, the environment or the formation scenario.
The slopes of the total density distribution peak at values of $\gamma_{\rm tot} \approx -2.1$, with a tendency to steeper slopes for less massive, more compact systems with lower dark matter fractions within the half-mass radius.
This is in agreement with observational results found for the Coma Cluster ellipticals \citep{thomas:2007MNRAS.382..657T,thomas:2009ApJ...691..770T} as well as from strong-lensing (\citealp{auger:2010ApJ...724..511A,barnabe:2011MNRAS.415.2215B,sonnenfeld:2012ApJ...752..163S}).

The cosmological simulations show a similar range of values in the density slopes as the observations in agreement with a similar analysis by \citet{lyskova:2012MNRAS.423.1813L} for the subset of cosmological zoom in simulations.
While our isolated binary mergers of present-day galaxies cannot reproduce the observed range of values, binary merger between two high-redshift spirals might be able to achieve better results since the initial conditions for those galaxies would look very different.
This provides evidence to the idea that, at least in a cluster environment like Coma, the elliptical galaxy population has not been formed by recent major mergers but rather at higher redshifts.
This is consistent with the fact that elliptical galaxies have higher dark matter densities than spiral galaxies which indicates a higher assembly redshift of about $z\sim 3$ (see for example \citealp{gerhard:2001AJ....121.1936G,thomas:2009ApJ...691..770T}).
Additionally, there is strong evidence that massive ellipticals formed a significant amount of their stars at high redshifts (for example \citealp{brinchmann:2000ApJ...536L..77B}), supported by observations of massive compact ellipticals at high redshifts (e.g., \citealp{vanDokkum:2009Natur.460..717V,vandeSande:2011ApJ...736L...9V}).

There is a small fraction of ellipticals from those observations that cannot be reproduced by our simulated elliptical sample.
These objects are characterized by small dark matter fractions, relatively flat total density slopes and large dynamical mass-to-light ratios in the case of the Coma Cluster ellipticals (see \citealp{thomas:2011MNRAS.415..545T}).
A possible explanation for these ellipticals could be that they have a bottom-heavy stellar IMF.
Non-universal stellar IMFs have been widely discussed recently, for example by \citet{cappellari:2012Natur.484..485C,ferreras:2013MNRAS.429L..15F,vanDokkum:2011ApJ...735L..13V,conroy:2012ApJ...760...71C,vanDokkum:2012ApJ...760...70V,treu:2010ApJ...709.1195T}, especially in the case of massive ellipticals.
The predicted dark matter fractions strongly depend on the assumed IMFs, and assuming for example a Kroupa IMF would result in dark matter fractions for the Coma Cluster ellipticals that are much higher and thus are in good agreement with our simulated results.
Nevertheless, as shown by \citet{conroy:2012ApJ...760...71C} and \citet{wegner:2012AJ....144...78W}, not even a variable stellar IMF can explain all high dynamical mass-to-light ratios that are observed.

We also see the effects of adiabatic contraction in case of the cosmological simulations:
While the density slopes for the dark matter component of these ellipticals peak around $\gamma_{\rm DM} \sim -1.67$, in good agreement with the results from \citet{sonnenfeld:2012ApJ...752..163S}, the dark matter only comparison sample peaks around $\gamma_\text{DM only} \sim -1.46$.
This is consistent with a number of other studies, for example \citet{jesseit:2002ApJ...571L..89J,blumenthal:1986ApJ...301...27B,gnedin:2004ApJ...616...16G,gnedin:2011arXiv1108.5736G} and especially \citet{johansson:2012ApJ...754..115J}.

The total density slopes correlate with the fraction of stars that are formed in situ: the steeper the slope, the larger the fraction of stars within the galaxy that were formed in situ and the lower the fraction of stars that were accreted.
This is in agreement with our result that the gas fraction in the binary merger scenarios is the only component that can significantly alter the slope of the total density distribution of the systems.
A higher star formation rate in the center of the newly formed elliptical due to a higher initial gas fraction causes a more prominent contribution from the new born stars to the total density and thus a steeper total density slope.

At higher redshifts, where gas and in situ star formation dominate the galaxies, the ellipticals from cosmological simulations have a total density slope of $\gamma_{\rm tot} \approx -3$, evolving through merger events toward a slope of $\gamma_{\rm tot} \approx -2$, supporting the idea of the two-phase formation of galaxies \citep{oser:2010ApJ...725.2312O}.
If the in situ fraction of a galaxy at $z=0$ is high, then the galaxy has accreted less stars from its environment in the second phase of its formation than a galaxy with a low in situ fraction at $z=0$.
Without enough gas-poor accretion in the second phase of formation, the total density slope could not changed a lot toward $\gamma_{\rm tot} = -2$, and thus the slope stays close to $\gamma_{\rm tot} \approx -3$.
This is in agreement with results presented in \citet{johansson:2012ApJ...754..115J} but in disagreement with observations from strong lensing by \citet{ruff:2011ApJ...727...96R} and \citet{bolton:2012ApJ...757...82B} who find a slight trend indicating that the slopes of the total density profiles of early-type galaxies at higher redshift are slightly flatter.
Nevertheless, this observed trend is very weak, and it could be due to a bias in the weak-lensing sample toward merging systems, as reported by \citet{torri:2004MNRAS.349..476T}, who found that merging systems tend to boost strong lensing. 
An enhancement in the lensing efficiency reported by \citet{zitrin:2013ApJ...762L..30Z} for an observed merging cluster of galaxies also supports this idea.
This would be in agreement with our result that merging systems show much flatter density slopes.

Our simple model predicts that the steepness of the slope of present-day galaxies is a measurement for the importance of the mergers the elliptical galaxy went through in its second formation phase.
Ellipticals with steep slopes close to $\gamma_{\rm tot} \approx -3$ had (nearly) no merger event in this second phase, while ellipticals with slopes around $\gamma_{\rm tot} \approx -2$ had a strong, collisionless merger dominated second formation phase.
Since our ellipticals with slopes around $\gamma_{\rm tot} \approx -2$ have generally higher dark matter fractions, this is consistent with the results presented by \citet{hilz:2012MNRAS.425.3119H,hilz:2013MNRAS.429.2924H} who show that the accretion of several small satellite systems strongly increases the dark matter fraction within the half-mass radius. 
From both, binary merger ellipticals and ellipticals from cosmological framework, we see that, once an elliptical has reached a total density slope of $\gamma_{\rm tot} \approx -2$, further merger events do not change the slope anymore.

We conclude that the density distributions with a slope of $\gamma_{\rm tot} \approx -2$ acts as attractors, independent of the individual stellar mass distributions of ellipticals.
We suggest that all elliptical galaxies will in time end up in such a configuration.
However, the mechanism that leads to this attractor is still unclear.
A possible explanation could be that this attractor state is a result of violent relaxation \citep{lyndenBell:1967MNRAS.136..101L}, although \citet{hilz:2012MNRAS.425.3119H} showed that violent relaxation is less efficient in minor merger events then in major merger events.

In general, the relaxation times for the elliptical galaxies from cosmological frameworks are much faster then for isolated binary mergers due to the presence of substructures and potential fluctuations.
We conclude that, to understand the entire range of elliptical galaxies observed at present day and their complex evolution scenarios, a full cosmological treatment is needed.

\begin{acknowledgements}
We thank Ortwin Gerhard, John Kormendy and Tomasso Treu for helpful discussions.
The authors also thank the anonymous referee for helpful suggestions and comments that improved this work.
The numerical simulations were performed on the local SGI-Altix 3700 Bx2, which was partly funded by the Cluster of Excellence: ``Origin and Structure of the Universe''.
R-S.R. acknowledges a grant from the International Max-Planck Research School of Astrophysics (IMPRS).
A.B. and T.N. acknowledge support from the DFG priority program SPP 1177.
P.H.J. acknowledges the support of the Research Funds of the University of Helsinki.
K.D. acknowledges the support by the DFG Priority Programme 1177 and additional support by the DFG Cluster of Excellence ``Origin and Structure of the Universe''. 

\end{acknowledgements}

\bibliography{bibliography}

\end{document}